\documentclass[iop,appendixfloats]{emulateapj}

\usepackage[usenames,dvipsnames]{color}
\usepackage{color}
\usepackage{amsmath}
\usepackage{float}
\usepackage{ctable}
\usepackage{graphicx}
\usepackage{refcount}
\usepackage{lineno,hyperref}
\usepackage[normalem]{ulem}
\newcommand{\figurepath}{.}

\newcommand{\cs}{c_{\rm s}}

\newbox\grsign \setbox\grsign=\hbox{$>$}
\newdimen\grdimen \grdimen=\ht\grsign
\newbox\laxbox \newbox\gaxbox
\setbox\gaxbox=\hbox{\raise.5ex\hbox{$>$}\llap
     {\lower.5ex\hbox{$\sim$}}}\ht1=\grdimen\dp1=0pt
\setbox\laxbox=\hbox{\raise.5ex\hbox{$<$}\llap
     {\lower.5ex\hbox{$\sim$}}}\ht2=\grdimen\dp2=0pt

\shorttitle{radiation map of the disk-jet}
\shortauthors{}
\begin{document}
\title{Dust continuum radiation maps from MHD simulations of accretion-ejection systems around single and binary stars}
\author{Somayeh Sheikhnezami
\altaffilmark{1} 
        Christian Fendt
\altaffilmark{2}
    Sareh Ataiee 
\altaffilmark{3}    
}
\altaffiltext{1}{Department of Physics, Institute for Advanced Studies in Basic Sciences (IASBS), Zanjan 45137-66731, Iran}
\altaffiltext{2}{Max Planck Institute for Astronomy, Heidelberg, Germany}
\altaffiltext{3}{Department of Physics, Faculty of Sciences, Ferdowsi University of Mashhad, Mashhad, 91775-1436, Iran}
\email{snezami@iasbs.ac.ir, fendt@mpia.de}  

     \date{\today}
\begin{abstract}

We study the launching of magnetized jets from a resistive circumstellar disk within a binary system, employing a unique 
combination of 3D MHD jet launching simulations (PLUTO code) and post-processed 3D radiative transfer modeling (RADMC-3D code).
Our findings reveal a well-defined jet originating from the inner region of the disk, extending to a larger disk area.
While the model attains steady states for a single star, 
a binary system leads to the emergence of tidal effects such as the formation of ``spiral arms'' in the disk and inside the jet. 
Here we have consistently implemented a time-dependent Roche potential for the gravity of the binary.
As a major step forward, we further present the first 3D radiation maps of the dust continuum for the disk-jet structure.
In principle, this allows us to compare MHD simulation results to observed disk-outflow features.
We, therefore, present convolved images of the dust continuum emission, employing exemplary point spread functions of the
MIRI instrument (5~$\mu m$ band) and the ALMA array  (320~$\mu m$ band).
In these bands, we identify distinguishable features of the disk-jet structure, such as "spiral arms," which we have also seen in the MHD dynamics.
 For gas density increased by an order of magnitude, the disk become optically thick at 5~$\mu m$, but remains bright at 320~$\mu$m.
At this wavelength, 320~$\mu$m, enhanced structural features in the disk and the base of the wind become more pronounced and are well resolved in the convolved image.
\end{abstract}


\section{Introduction}
It is now well known that one of the important phases of star formation is the formation and launch of stellar jets, 
which are high-velocity, collimated outflows of material ejected by a circumstellar disk.
The stellar jets are spectacular phenomena and play a crucial role in regulating star formation by removing excess angular
momentum and facilitating mass accretion onto the protostar.

There is a vast body of literature focusing on the study of jet formation through numerical simulations. 
Among these, pioneering work has been conducted by researchers such as \citet{1985PASJ...37...31S,1995ApJ...439L..39U,1997ApJ...482..712O}, 
following the earlier seminal analytical approaches by \citet{1982MNRAS.199..883B,1983ApJ...274..677P,1985PASJ...37..515U, Uchida1985}. 
These simulations explore the formation of jets from the disk surface, considering the acceleration of jet material and its collimation by 
the magnetic field \citep{1993ApJ...410..218W,1995ApJ...444..848L,1997A&A...319..340F,2002A&A...395.1045F,2010ApJ...709.1100P,2011ApJ...742...56V,2024ApJ...966...82S}.

However, it is essential to include the disk physics in the numerical treatment to understand the jet launching process 
from the inner part of the underlying disk.
Today, numerical simulations of the accretion-ejection process play an essential role in understanding jet launching.
A vast amount of literature exists on magnetohydrodynamics (MHD) simulation on jet launching, with ever-improving physical complexity 
and also numerical resolution
\citep{Uchida1985, 1998ApJ...508..186K, 2002ApJ...581..988C, 2007A&A...469..811Z,2010A&A...512A..82M, 
2012ApJ...757...65S, 2014ApJ...793...31S, 2018ApJ...861...11S}.
In general, these works study how the properties of the outflow that is formed from the disk are determined from
certain disk properties, namely the disk resistivity, the presence of the mean-field dynamo in the disk, or a 3D circumstellar disk in a Roche potential.

Recently, detailed discussions have taken place regarding the full 3D simulations of the disk-jet 
structure in both single stars and binary 
star systems using rectangular coordinates \cite{2015ApJ...814..113S,2018ApJ...861...11S,2022ApJ...925..161S}.
However, to optimize computational efficiency, it is worthwhile to develop the model from rectangular
to spherical coordinates.

It is well known that protostellar disks undergo accretion. 
However, the exact mechanism responsible for extracting angular momentum and driving accretion in the low-ionization {"}dead{"} zones of
the disk is a matter of debate.
 
 In recent years, magnetohydrodynamic (MHD) disk winds have emerged as a leading candidate for this process.
 The study of accretion and ejection has gained significant attention in the context of protoplanetary disks, with recent research focusing 
 on the role of MHD disk winds during the late stages of protostellar or protoplanetary disk evolution.
For instance, \citet{2022MNRAS.512L..74T} show that the essential features of disk evolution as probed by the observations are naturally reproduced by MHD wind-driven accretion. 
They show that MHD disk winds alone can account for both disk dispersal and accretion properties and how MHD winds can explain the correlation between disk masses and accretion rates in Lupus.
In addition, \cite{2022A&A...668A..78D} gave evidences for MHD wind in class~I object DG~Tauri~B. 
Their results provide the strongest evidence so far for the presence of massive MHD disk winds in Class I sources with residual infall, and they suggest that the initial stages of planet formation take place in a highly dynamic environment.

Interestingly, \cite{2024A&A...686A.201N} present high angular resolution (30 AU) ALMA observations of the emblematic L1448-mm protostellar system and 
find suggestive evidence for an MHD disk wind. Various system components (i.e., disk, inner envelope, and rotating outflow) are resolved for the first time in their observation.
Further literature examines the role of MHD disk winds in protostellar and protoplanetary disks. For instance, \citet{2023ASPC..534..567P} provide a review on disk winds in protoplanetary disks, and \citet{2023ApJ...954L..13S} discuss methods to disentangle the effects of disk winds and viscosity in disk observations.

In this paper, we first introduce our 3D model setup, which involves a circumstellar disk launching a magnetized jet in the presence 
of resistivity. 
Spherical coordinates are used to describe the system accurately with high resolution.
We have implemented a companion star into the model setup and follow the full 3D evolution of the disk-jet in the single star 
and binary star systems.
This allows us to investigate the companion star's influence on the jet and disk dynamics.

The primary goal of our study is to explore the dynamical evolution of the accretion-ejection structure in these different systems. 
In addition, {\em as a major step forward, we will present the first 3D radiation map of dust continuum for the disk-jet structures}, 
utilizing the radiative transfer code {"}RADMC-3D{"}.
In principle, this can allow us to compare MHD simulation results to observed disk-outflow features.

 Our paper is organized as follows: 
 In Section 2, we introduce our model setup.
 In section 3, we present and discuss the dynamics in our 3D simulation of a disk-jet launched from a single star and a binary star system. 
 In section 4, we discuss the radiative transfer model we applied to obtain the radiation map of the disk-jet.
 In section 5, we present the dust continuum radiation map obtained for the disk-jet structure. 
 In Section 6, we summarize the results.

\section{Model setup}
In this paper, we study the formation of the magnetized jet from a circumstellar disk staying in a single star and investigate the radiative 
signatures of dust particles present in the disk and the jet.
In the following, we describe our dynamical MHD modeling utilizing the PLUTO code \citep{2007ApJS..170..228M,2012ApJS..198....7M}, 
and the radiative transfer on the simulated MHD data applying the code 
RADMC3D \citep{2012ascl.soft02015D}\footnote{{\tt http://www.ita.uni-heidelberg.de/}$\sim${\tt dullemond/\\software/radmc-3d}}.

\begin{table}
\caption{Characteristic simulation parameters, as chosen input parameters for our simulation runs.
We show the plasma beta $ \beta_p= 8 \pi P / B^2$, 
the binary separation $D$, 
the orbital period of the binary $P_{\rm b}$, the disk temperature at inner radius $T_{\rm i}$ , the mass ratio of the secondary to the primary star $q= M_s / M_p$ and  the outer radius of the computational domain $r_{out}$.
All simulations assume a plasma-beta $\beta_{p}= 1000$.
The physical parameters are measured at the inner disk radius at the disk mid-plane.
  The smaller separation chosen for the binary run allows for a faster evolution of the system to capture the 3D effects.
}
\begin{center}
\begin{tabular}{c|c|c|c|c|c|c}
\hline
\hline
\noalign{\smallskip}

Run  &  D  & $q$ & $P_{\rm b}$ & $\rho_{\rm du}/\rho_{\rm gas}$ & $T_{\rm i}$ & $r_{out}$ \\
\noalign{\smallskip}
\hline
\noalign{\smallskip}
\noalign{\smallskip}
{\em rsin}    & 0    &  0   & -     & 0.01 & 600-1000 & 1000 $r_{\rm i}$ \\
 {\em rb1} & 300  & 0.01 & 29803 & 0.01 & 600-1000 &280 $r_{\rm i}$ \\
{\em rb2}     & 300  & 0.2  & 32486 & 0.01 & 600-1000 &280 $r_{\rm i}$ \\
\noalign{\smallskip}
 \hline
 \noalign{\smallskip}
 \end{tabular}
 \end{center}
\label{tbl:0}
\end{table}

\subsection{Dynamical modeling}
We first need to establish a proper model setup for a jet formed in a single star system.
We then implement the Roche potential and consider a binary system with a {"}primary{"} of mass $M_{\rm p}$ and a {"}secondary{"} of mass $M_{\rm s}$, 
separated by the distance $D$.
The origin of the coordinate system stays in the primary, and the location of the secondary is chosen to be outside the computational domain (see Figure \ref{com_dom}).

We then derive radiation maps of the dust continuum for both the single star-disk-jet and binary star-disk-jet systems. 
Finally, we showcase enhanced radiation maps that have been refined by incorporating convolution and noise from a specified telescope (such as ALMA) to provide insights into the expected radiation map characteristics of the aforementioned systems.

\subsubsection{Governing equations}
In the current paper, we present the results of simulations for which we 
have applied the MHD code PLUTO
\citep{2007ApJS..170..228M, 2012ApJS..198....7M}, version 4.4.2,
to solve the time-dependent, resistive, inviscid MHD equations,
accounting namely for the conservation of mass, momentum, and energy,
\begin{equation}
\frac{\partial\rho}{\partial t} + \nabla \cdot \left( \rho \vec u \right)=0,
\label{continuity}
\end{equation}
\begin{equation}
\frac{\partial \left( \rho \vec u \right) } {\partial t} + 
\nabla \cdot \left(  \rho \vec u \vec u \right) + \nabla P-\frac{ \left( \nabla \times \vec B \right) \times \vec B}{4 \pi}
+ \rho \nabla \Phi = 0.
\label{momentum_eq}
\end{equation}

\begin{multline}
 \frac{\partial e}{\partial t} + \nabla \cdot \left[ \left( e + P + \frac{B^2}{8\pi} \right) \vec u - \left( \vec u \cdot \vec B \right) \frac{\vec B}{4\pi}  \right] = 
  \Gamma \nabla \cdot  \left[ {\eta} \vec j \times \frac{\vec B}{4\pi} \right].   
\end{multline}

Here, $\rho$ is the mass density, $\vec u$ is the velocity, $P$ is the thermal gas pressure,
$\vec B$ stands for the magnetic field, and $\Phi$ denotes the gravitational potential.

In the above equations, $\eta$ denotes the magnetic diffusivity (or resistivity, respectively).
We consider an anomalous magnetic diffusivity that is turbulent in nature and, thus, 
cannot be derived self-consistently in our setup.
Since the disk is resistive, ohmic heating will occur (see right hand side of energy equation).
For simplicity, here, we neglect ohmic heating by setting the parameter $\Gamma=0$, assuming that ohmic heating energy 
is radiated away immediately - corresponding to a hypothetical cooling term that equals to the ohmic heating term.
This makes indeed sense as we do not consider radiation transport in our MHD approach, which would thus lead to an 
arbitrarily large heating and thus disk flaring.

We adopt an $\alpha$-prescription \citep{1973shakuraetal}, consistent with the assumption of a turbulent 
origin, as supported by previous studies \citep{2007A&A...469..811Z,2012ApJ...757...65S,2014ApJ...793...31S}.
The disk diffusivity profile is defined by 
$\eta(r,z) = \alpha u_{A}(r,z=0,t) H(r) F_{z,H}$, 
where the Alfv\'en velocity $u_A$ and the disk thermal scale height $H(r) = c_{\rm s}(r,z=0)/\Omega_K(r,z=0)$ 
are evaluated at the disk mid-plane.
 Here, $\alpha$ is the magnetic diffusivity parameter and its value is 1.6 in our model.
The function $F(z,H)$ further confines the diffusivity to the disk region and up to one-two scale height above the disk surface, and it is defined as 

\vspace{0.2cm}
$ F{(z,H(r))} = \left\{
    \begin{array}{lll}
       \exp\left(-0.5\left(\frac{z-H(r)}{H(r)}\right)^{2} \right) & &  z>H(r)\\
       1                            &  &   z\le H(r)\\
    \end{array}
\right.$
\vspace{0.3cm}

So far, we have considered the (turbulent) resistivity as the only diffusion term for the disk. 
However, additional diffusion, such as ambipolar diffusion, arising from the drift between charged and neutral particles in a 
partially ionized plasma, also exists in the disk, in particular for large radii.
The coupling between gas, dust and the magnetic field could in principle affected by this effect, 
correspondingly changing accretion and outflow rates.
However, to fully understand its impact, the ambipolar timescale must be considered in relation to the disk or jet dynamical timescales 
(which has been studied e.g. by \cite{1999ApJ...524..947F}).
While the effect may play a role for the disk and jet dynamics, the treatment of ambipolar diffusion is beyond the scope of this paper.

The electric current density $\vec j$ is given by Amp\'ere's law 
$\vec j = \left( \nabla \times \vec B \right) / 4\pi$, while 
the total energy density is
\begin{equation}
e = \frac{P}{\gamma - 1} + \frac{\rho u^2}{2} + \frac{B^2}{2} + \rho \Phi.
\end{equation}
with an ideal gas with a polytropic equation of state $P = (\gamma - 1) u$ with the polytropic index of $\gamma = 5/3$ and the internal energy density $u$.

The evolution of the magnetic field is described by the induction equation,
\begin{equation}
\frac{\partial \vec B}{\partial t} - \nabla\times \left( \vec u \times \vec B - \eta \vec j \right) = 0.
\end{equation}

\subsection{Gravitational potential}
We consider a time-dependent gravitational (Roche) potential $\Phi= \Phi_{\rm eff}$. 
Since the origin of our coordinate system is in the primary, we have to consider the time variation of the gravitational potential
in that coordinate system.
We prescribe the position of the secondary initially $(t=0)$ along the $x$-axis.
Thus, its position vector varies over time (in Cartesian coordinates) as,
\begin{equation}
\vec{D}= \hat{x} D \cos{\omega t} + 
         \hat{y} D \sin{\omega t}\cos{\delta} + 
         \hat{z} D \sin{\omega t}\sin{\delta},
         \label{roche_potential}
\end{equation}
Here, $\delta$ denotes the inclination angle of the binary orbit with respect to the initial circum primary disk.
In addition, $\hat{x}$, $\hat{y}$ and $\hat{z}$ represent the unit vectors in Cartesian coordinates. 
In this paper, we discuss a co-planar geometry and thus $\delta = 0$.
The effective potential in a binary system at a point with position vector $\vec{R}( x, y, z )$ is
\begin{equation}
\Phi_{\rm eff} = - \frac{G M_{\rm p}}{|\vec R|} - \frac{G M_{\rm s}}{|\vec{R}-\vec{D}|}
                 + \frac{G M_{\rm s}}{|\vec{D}|^3}  \left(\vec{R} \cdot \vec{D}\right).
\label{eq:phi_eff}
\end{equation}
The terms in Equation~\ref{eq:phi_eff} describe the gravitational potential of the primary, followed by terms describing 
the tidal perturbations caused by the orbiting secondary. 
The last "indirect" term accounts for the acceleration of the origin of the coordinate system
(see also \citealt{1996MNRAS.282..597L,2018ApJ...861...11S}).

\begin{figure}
\centering
\includegraphics[width=1.\columnwidth]{\figurepath/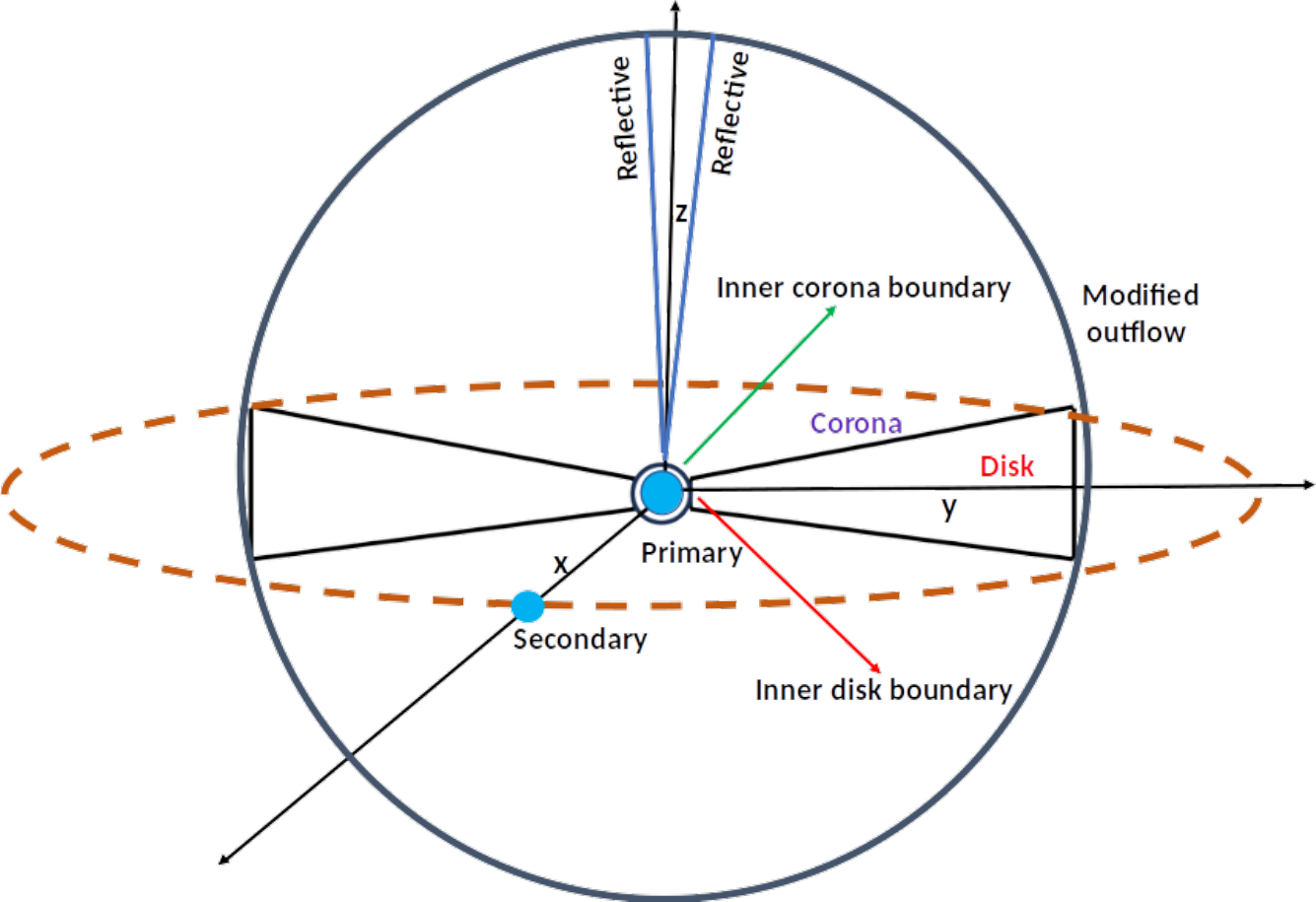}
\caption{Computational domain. 
Shown is a schematic view of the 3D spherical computational domain that we use for the the presented simulation in this paper. 
The origin of the coordinate system is at the center of the primary (blue). 
The secondary (blue) is orbiting (dashed red line) outside the outer boundary (outer black circle (sphere)).
The inner boundary (small black circle) is located at the (initial) inner disk radius.
The black circles indicate spheres in our 3D spherical coordinate system.}   
\smallskip
\label{com_dom}
\end{figure}

\subsection{Units and normalization}
\label{units}
Here, we describe the normalization of units applied consistently throughout this paper. 
The inner radius of the disk boundary is designated as $r_{\rm i}$, with a normalized value of $r_{\rm i}=1$. All measurements are expressed in units relative to this inner disk radius.

We want to stress that the astrophysical scaling of the variables in our simulations is
predominantly motivated by previous numerical studies. 
We are not yet in a position to fit certain sources, as many parameters for theses sources still remain unknown.
For example, any {\em "observed"} accretion rate depends on an underlying model. 
That model, however, will be different from our modeling.
On the other hand, real accretion rates we expect to be derived at some point from time-dependent simulations
including a thorough treatment of radiation.

However, in general, our normalization can be understood as applicable to a relatively evolved young low-mass star, such as a jet-emitting classic T\,Tauri that is not embedded anymore (class I or II object).

We normalize all variables, namely $r, \rho, v, B$ to their fiducial values at the inner disk radius, $r_{\rm i}$.
A change in one of the system parameters accordingly changes the scaling of our simulation results
which can thus be applied to a variety of jet sources.

The gas pressure and density at the inner disk radius $r_{\rm i}$ are denoted as $P_{\rm i}$ and $\rho_{\rm i}$, respectively.
Velocities are scaled according to the Keplerian velocity at the inner radius of the disk, denoted as $v_{\rm k, i}$, 
which is normalized to $v_{\rm {k,i} = 1}$.
Time units are defined in relation to $t_{\rm i} = {r_{\rm i}}/{v_{\rm k,i}}$.

An important consideration is the astrophysical scaling of the inner disk radius. 
Since our simulation results are expressed in code units, the inner disk radius can be specified rather arbitrarily, 
depending on the choice of the astrophysical system we want to consider. 
Since our study focuses on the spiral arm structure which is predominantly located in the outer part of the disk,
we have adopted $r_{\rm i} = 1$ AU as our inner disk radius, which we think is an appropriate choice.
Moreover, the choice of $r_{\rm i} = 1$ AU results in a cooler disk temperature at the inner disk.
Consequently, dust particles will be found also in the inner disk of our simulation.
We note that we cut out the very inner, hotter disk from $0.1 - 1.0$ AU from our treatment.

Furthermore, 
using a larger disc, we can provide the radiation map for the outer part of the disk where is more suitable to observe by the currently available instruments.

The normalization of density remains a choice.
In pure MHD simulations we are treating a Keplerian problem, that in principle can constrain the central mass, but
not the mass of the orbiting body.
If radiative transfer would be included, the opacity and radiation fluxes derived from the simulation would allow for a comparison with observations, and thus provide a constrain on the disk density.
However, obviously further parameters enter the problem, as e.g. the chemistry of the disk.

For the present paper, we decided to stay close to previous simulation work with the density normalization (see discussions
in e.g. \citet{2012ApJ...757...65S,2014ApJ...793...31S}).
We choose $\rho_{\rm i}=10^{-12} \rm gr \,cm^{-3}$ as the equatorial disk density at the inner disk radius, generally.
However, for comparison we will also offer radiation map results applying higher disk densitiy, $\rho_{\rm i}=10^{-11} \rm gr \,cm^{-3}$.
These models may be applicable to younger jet-driving protostars.

The mass accretion rate is, in principle, an observable, however, the derivation of a numerical value from observations is model-dependent.
Depending on the disk model utilized, the observed disk luminosity can be correlated to
an accretion rate.

The accretion rates are typically expressed in units of 
$\dot{M}{\rm i} = {r_{\rm i}}^2 \rho_{\rm i} v_{\rm K,i}$ and in young stellar object (YSO) models are about $ 10^{-7} {\rm M{\odot} , yr^{-1}}$  
for a density of $\rho_{\rm i}=10^{-12} {\rm gr}\,{\rm cm}^{-3}$ and an inner radius of $0.1 AU$ \citep{2007A&A...469..811Z,2012ApJ...757...65S}.
Clearly, accretion rates may vary significantly between individual sources and also evolve over time.

While the assumed disk density will govern the derived disk accretion rates  $\dot M_{\rm i} \propto \int \rho v_R ds$,
the radiation fluxes derived from the treatment of dust particles will depend on another {"}free{"} parameter, that is
the dust to gas ratio - and obviously the dust properties. 
We will discuss this further below in Section~\ref{Radiation modeling}.

The magnetic field strength we measure in units of the magnetic field at the inner disk radius, represented as 
$B_{\rm i} = \sqrt{ 2 P_{\rm i} / \beta_{\rm i} }.$
Here, ${\beta_{\rm p}}$ represents the plasma beta parameter of the gas.\footnote{In PLUTO code the magnetic field 
is normalized considering $4\pi \equiv 1$ 
\label{MyFootNoteLabel}}.

To our understanding, the disk magnetic field strength is basically un-known.
MHD simulations show that a certain field strength is needed to launch powerful outflows 
(see e.g. \citet{2012ApJ...757...65S, 2014ApJ...793...31S}).
Thus, again, we choose a field strength that is similar to previous numerical studies, as we want to examine binary disks that
do launch outflows.
With our choice of parameters, a plasma-beta $\beta_i \simeq 1000$,

we consider a field strength $(B_{\rm i} \simeq 14.9\,$Gauss.
The choice of field strength will not directly affect the radiation pattern.
But as it governs the strength of the jet, its mass flux and density (and thus its dust loading),
so it impacts the jet radiation pattern indirectly.

The temperature profile of the disk is obtained from the polytropic gas law and will be discussed extensively in section~\ref{Radiation modeling}.

\subsection{Numerical setup}
We employ spherical coordinates $(R, \theta, \phi)$ for our simulations (see Figure~\ref{com_dom}).
The computational domain extends from 0 to $1000$ code units of length $r_{\rm i}$ in the radial direction.
In the polar direction, the domain covers a range from 0 to $\pi$ radians,  with a small opening angle near the polar axis.

\begin{figure}
\centering
\includegraphics[width=1.\columnwidth]{\figurepath/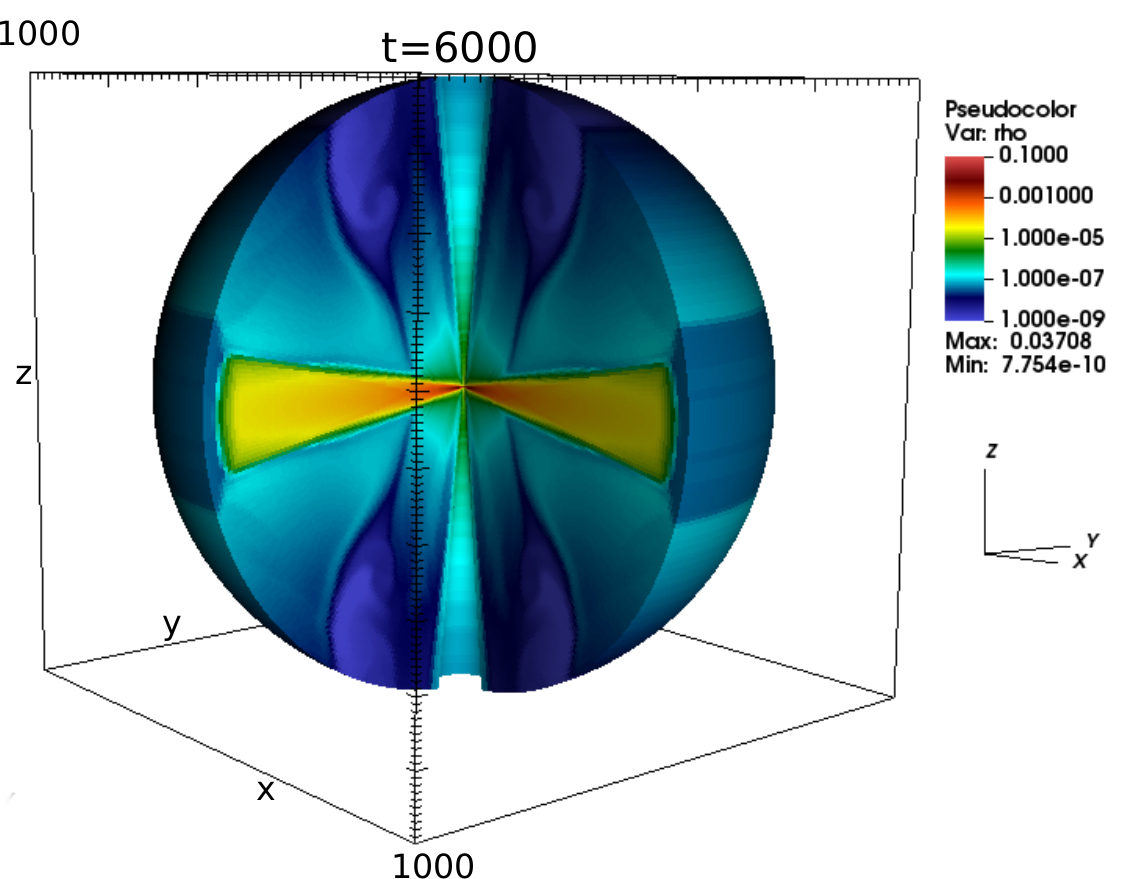}
\caption{Gas density. Shown is the 3D rendering of the mass density of the gas obtained from our simulations for the jet formed in a single star system from a circumstellar disk.}
\label{fig:3Dview_rho_sing}
\end{figure}

Our grid extends 1000$r_i$ in the radial direction, $\pi$ radians in the polar direction, 
and $2\pi$ radians in the azimuthal direction, including the total of 
$560\times192\times256$ cells.
In the radial direction, the cells are stretched, resolving the innermost part 
with the highest resolution of 0.04, but it increases for larger radii.
In both the polar and azimuthal directions, the cells are uniformly spaced, with a 
resolution of 0.0153 and 0.02454, respectively. 

In our simulations considering binary star-disk-jet systems, we apply a smaller computational domain in the radial direction, just to enhance potential tidal effects on the disk and jet.
By applying a smaller stellar separation between the primary and secondary stars, we also place the secondary star outside
the computational domain.
This results in a radial extent of 280$r_i$, incorporating 256 stretched grid cells.

\subsection{Boundary conditions}
The boundary condition in the azimuthal direction is set to be periodic. 
For the polar axis, we apply a reflective boundary condition, incorporating a small opening 
angle to achieve a better time step.
 At the outer boundary, we apply modified outflow boundary conditions, meaning that variables (i) are simply copied 
to the ghost cells (original PLUTO setup), but (ii) enforce in addition that matter is not allowed to enter the domain.

At the inner radius $r=1$, we define a specific condition, considering both the accretion flow
from the inner disk and also the coronal region from the disk surface to the rotational axis.

More specifically, in the disk region ($\pi/2-2\epsilon \le \theta \le \pi/2 +2\epsilon$) 
where $\epsilon$ represents the disk scale height, we apply accretion boundary conditions. 
Here, pressure and density follow a power-law profile in the radial direction, adopting the approach developed by \citet{2014ApJ...793...31S}, which is based on a self-similar gas expansion.
An {\em accretion condition} is applied for the radial velocity, while the polar 
velocity component is copied. 
The azimuthal velocity is determined using angular momentum conservation, 
and the toroidal magnetic field is constrained by electric current conservation. 
The other two components of the magnetic field are copied to the ghost cells.

For the corona region at the inner boundary, we keep the initial coronal density and pressure. 
In addition, we impose the condition $E_{\phi}=0$ from ideal MHD, ensuring $v_p || B_p$,
to define the radial and polar velocity components.
 Along the inner coronal boundary, we prescribe a weak inflow into the domain with $v_{\rm p} = 0.02$.
The azimuthal components of velocity and magnetic field are set to vanish to achieve a more stable structure.

\begin{figure*}
\centering
\includegraphics[width=18cm]{\figurepath/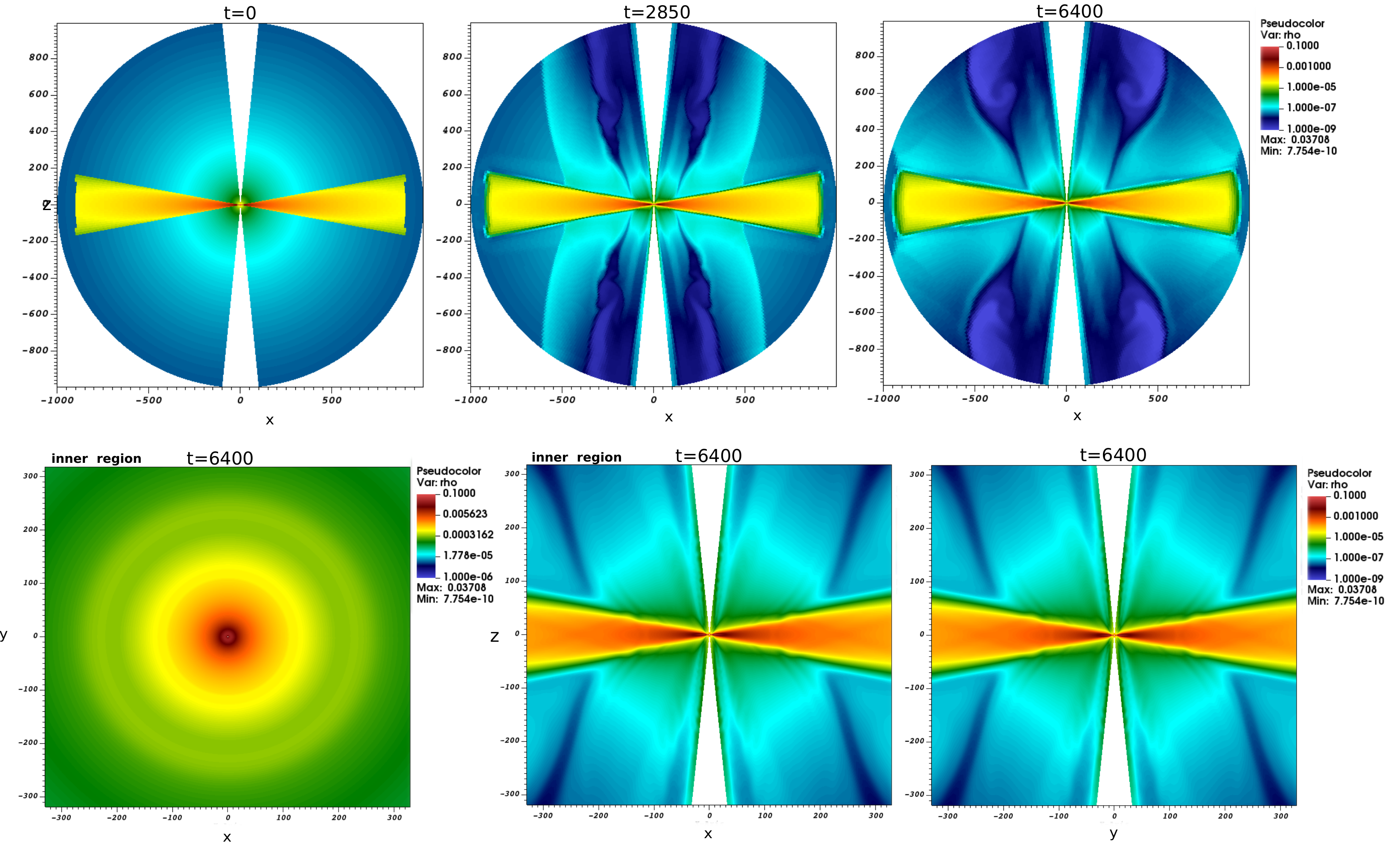}
\caption{Density distribution. 
 Shown are snapshots of the mass density of a jet formed from a circumstellar disk in a single star system at times $t=0, 2850, 6400$. 
 The top row shows slices in the $x$-$z$ meridional plane of the whole domain. 
 The bottom row displays differently directed slices of the gas density at the inner part of the disk-jet structure. }
\label{fig:2Dview_rho_sing}
\end{figure*}

\begin{figure*}
\centering
\includegraphics[width=18cm]{\figurepath/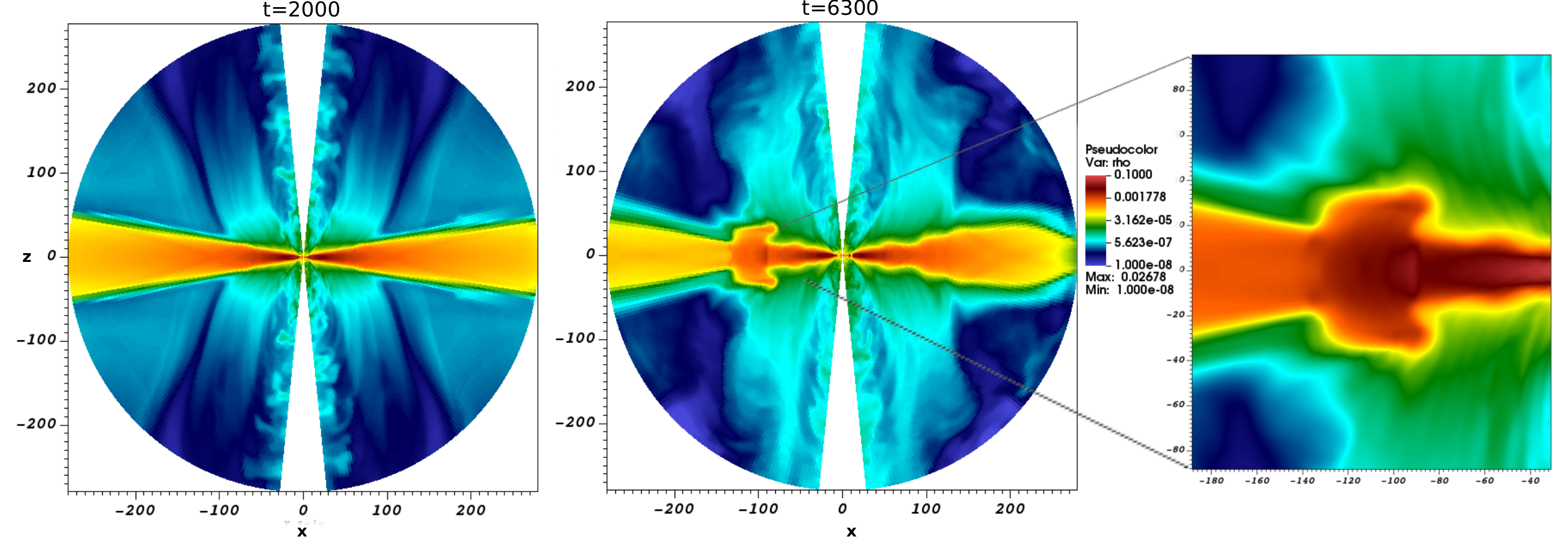}
\centering
\caption{Gas density. Shown are the 2D snapshots of the mass density of the gas obtained from our simulations for the jet formed in a binary star system with the mass ratio of $q=0.2$ of the secondary to primary star,  located at a distance of $D=300$ from the primary star, at $x-z$ plane.}
\label{fig:3Dview_rho_rb2_q0.2}
\centering
\includegraphics[width=18cm]{\figurepath/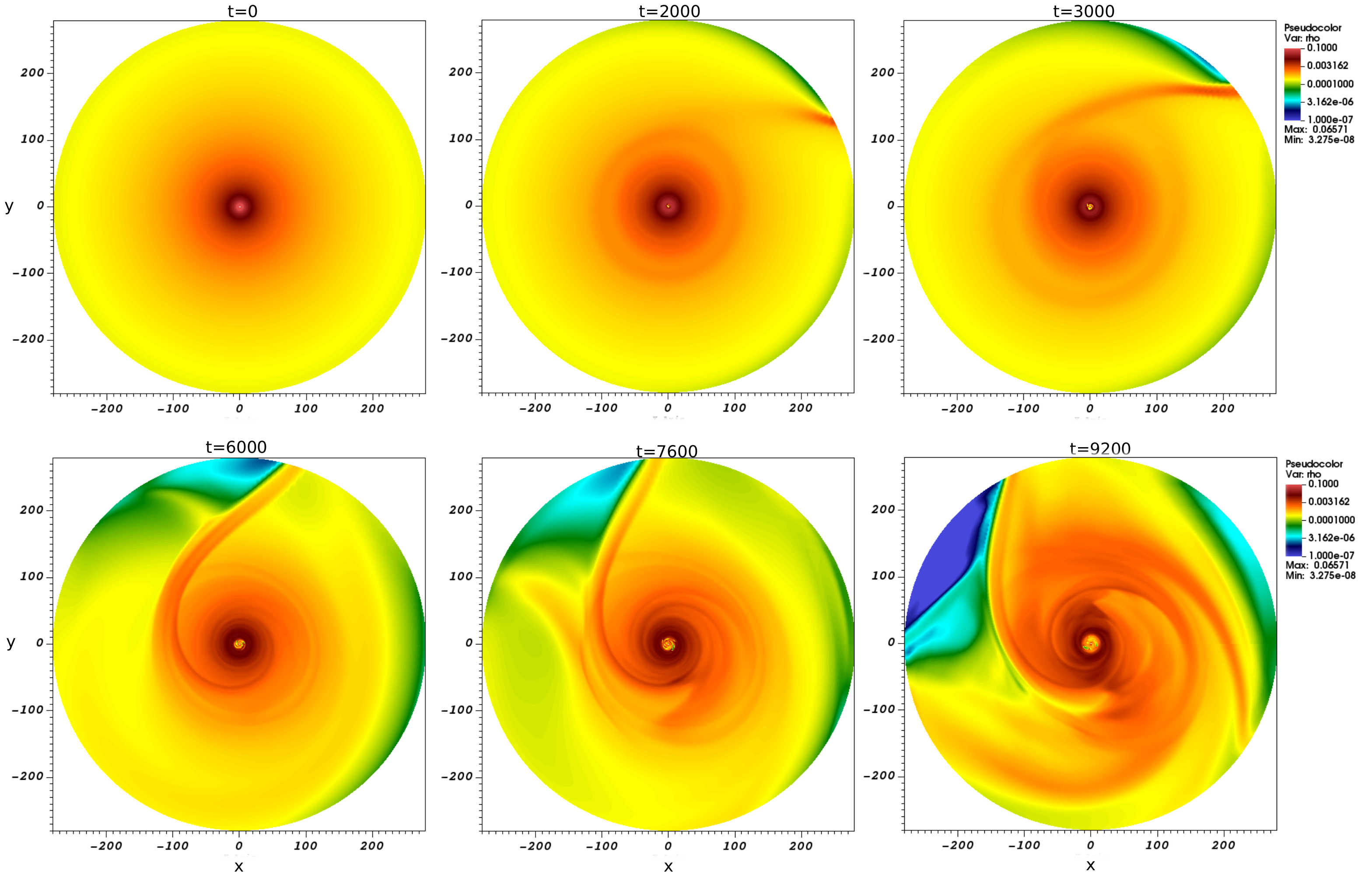}
\centering
\caption{Gas density. Shown are the snapshots of the mass density of the gas obtained from our simulations for the jet formed in a binary star system with the mass ratio of $q=0.2$ of the secondary to primary star,  located at a distance of $D=300$ from the primary star.
The snapshots correspond to different dynamical times at the disk midplane.}
\label{fig:2Dview_rho_rb2_q0.2}
\end{figure*}

\subsection{Initial conditions}
We adopt initial conditions consistent with those employed in previous studies, wherein a magnetically diffusive accretion disk is prescribed in sub-Keplerian rotation. 

Above the disk, a hydrostatic corona in pressure equilibrium with the disk is established \citep{2007A&A...469..811Z,2010A&A...512A..82M,2012ApJ...757...65S}. The coronal density is set to be several orders of magnitude lower than the disk density, resulting in a sharp entropy and density discontinuity at the disk-corona interface.
We prescribe an initially geometrically thin disk with
$\epsilon = H/r =  0.1$ which is itself in vertical equilibrium between thermal pressure and gravity of the primary star. The initial conditions, including the density and pressure profiles for the disk and corona, are adopted from our previous works\citep{2012ApJ...757...65S,2018ApJ...861...11S}. 

\subsubsection{Magnetic field distribution}

Following previous works \citep{2007A&A...469..811Z,2012ApJ...757...65S,2014ApJ...793...31S}, our simulations are initialized with a poloidal magnetic field, defined through the vector potential
$\vec B = \vec\nabla \times \vec A_{\phi}$ and is given by
\begin{equation}
\displaystyle
A_{\phi} = \frac{4}{3} B_{z,0} r_{\rm i}
            \left(\frac{r}{r_{\rm i}}\right)^{-1/4}
            \frac{m^{5/4}}{\left( m^2 + {\cot^2 \theta}\right)^{5/8}}.
\label{eq;magpsi}
\end{equation}
Here $B_{z,0}=\epsilon \sqrt{2/\beta}$ measures the vertical field strength at the inner disk radius and at mid-plane,
$r=r_{\rm i}$ and $z=0$.

\section{Dynamics - simulation results}
In this section, we present and discuss the results derived from our 3D MHD simulation of jet launching from a diffusive magnetized circumstellar disk for both a single star-disk-jet  and binary star-disk-jet system.
 
The primary objective of our study is to investigate the dynamical evolution of these systems.
We will later present the first 3D radiation map of the dust continuum for the disk-jet structures.

\subsection{A 3D disk-jet from a single star}
We present, for the first time, the results of a 3D MHD simulation of the disk-jet structure using spherical coordinates.
Two simulations are presented, as listed in Table \ref{tbl:0}. 
The first simulation corresponds to a jet launched from a circumstellar disk in a single star system. 
The simulation was run for approximately, 10000 time units $t_{\rm i}$, corresponding to 1600 revolutions of the inner disk at $r=r_{\rm i}$.

Figure~\ref{fig:3Dview_rho_sing} illustrates the 3D rendering of the gas mass density for the jet formed in a single star system, from a circumstellar disk.
Furthermore, Figure~\ref{fig:2Dview_rho_sing} showcases snapshots of the mass density in the $x$-$z$  plane.
The jet material is initially launched from the inner disk and, over time, expands to larger radii, thereby accelerating and collimating.

The simulation domain is large, extending to $1000 r_i$, which allows us to observe 
two distinct outflow components: the faster outflow or jet, originating from the inner disk
area, and the extended disk wind that is launched from larger disk radii.
The whole 3D structure shows perfect symmetry in all variables exhibited, both horizontally (azimuthally) and between the two hemispheres.

This simulation serves as a quality test for both the code and our model setup. 
In particular, the definition of a proper axial boundary condition is crucial for ensuring 
a symmetric structure, or, respectively, confirming that non-axisymmetric features are 
physical and not numerical.
In addition, without a small opening cone along the axial boundary, the time step may become
arbitrarily small.
Since this paper primarily focuses on the disk evolution influenced by tidal effects, 
incorporating a small opening angle does not significantly impact disk evolution, but it 
allows for simulating over a sufficiently long period.

The simulations presented in this paper confirm our previous findings \citep{2015ApJ...814..113S,2018ApJ...861...11S,2022ApJ...925..161S},
with the advantage of achieving a higher resolution and improved computational efficiency.

Also, these new simulations extend the 3D simulation work by \citet{2003ApJ...582..292O} by now including the disk evolution. 
There have been other studies on accretion disks using 3D simulations 
\citep{1996ApJ...463..656S, 2008MNRAS.386.1274L, 2011ApJ...735..122F, 2013MNRAS.430..699R}, 
however, our work is the first simulation of the jet launching from a circumstellar disk applying 3D {\em spherical coordinates} with respective high resolution.

\subsection{A 3D disk-jet from a binary system}
In this section, we present and discuss the results of another simulation in which 
we have implemented a companion star orbiting outside the computational domain. 
As a result, when the jet reaches larger distances from the launching site,
its structure and evolution are more affected by the (time-variable) 3D Roche potential.

In this work, we achieved higher resolution in our simulations while making them more efficient and resource-saving.
Additionally, when studying the disk, spherical coordinates with a curved azimuthal component are significantly more effective and 
consistent with the disk's shape than a rectangular coordinate system.

In Figure~\ref{fig:3Dview_rho_rb2_q0.2} we display the 2D snapshots of the gas density at $x-z$ plane as obtained from our simulation {\em rb2} that 
considers a jet formed in a binary star system with the mass ratio of $q=0.2$ and a binary separation $D=300 r_i$.
In Figure~\ref{fig:2Dview_rho_rb2_q0.2} we present snapshots of 2D slices of the mass density at the disk midplane.

Both figures demonstrate that as the disk evolves, axial symmetry is broken, and spiral arms are formed.

The density map clearly illustrates the growth and expansion of the spiral arms. 
In the absence of disk evolution, the direct tidal effect appears as an arc-like feature within the jet, as demonstrated by \citet{2024ApJ...966...82S}.

Our previous studies \citep{2018ApJ...861...11S,2022ApJ...925..161S} have shown that the spiral arms exhibit rotational motion synchronized
with the orbital motion of the secondary. 
This behavior is also observed in the current simulations. 
Another intriguing finding that supports our previous work is that the spiral formation is also evident in the outflow, which has been 
influenced by disk evolution and delivered into the jet.
 
Moreover, it is important to note that since the secondary star is positioned outside the computational domain, the influence of the disk 
(or back-reaction) on the companion star is not considered.

Additionally, some snapshots of the gas density from another binary run (rb1) with a mass ratio of $q = 0.01$ are shown in Figure~\ref{fig:3Dview_rho_rb1_q0.01}(in the appendix).
It is observed that while all similar tidal effects are present in this run, they are less pronounced due to the smaller mass ratio.

\begin{figure}
\centering
\includegraphics[width=0.7\columnwidth]{\figurepath/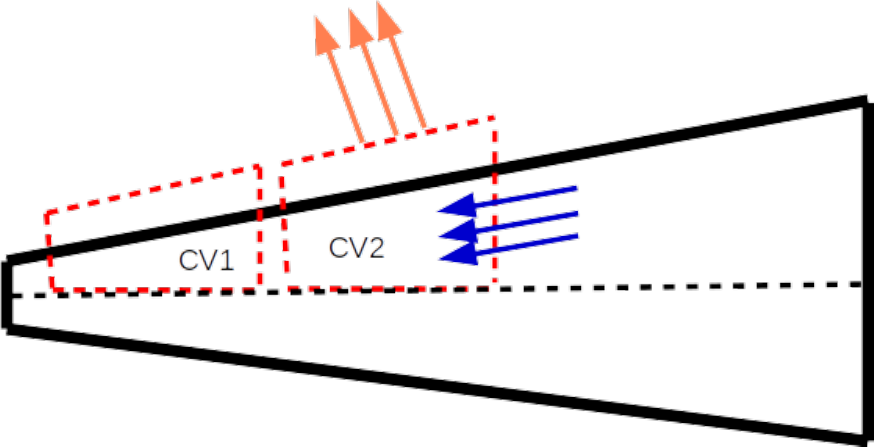}
\caption{Control volumes for integrating (and comparing) the accretion mass fluxes (blue) and ejection mass fluxes (red).}
\label{control_volume}
\end{figure}

\subsection{Mass fluxes in accretion and ejection}
One useful observable is the mass flux of the accreting mass and the outflow launched from the disk, which helps to determine the fraction of accreting material delivered
into the outflow.
It also provides insights into the evolution of the disk and the significance of the processes involved in driving the accretion and ejection processes.

Essentially, all the features and physical behavior observed in the outflow originate from the physical evolution of the underlying disk. 
In other words, the physical evolution of the ejected material is induced by the accreted material evolution.
This correlation can be analyzed by examining the fluxes of accreting and outflowing mass.

We measure the mass fluxes integrating two arbitrary defined control volumes (see Figure~\ref{control_volume}). 
The first control volume CV1, spans a radial range from $2\,r_{\rm i}$ to $12\,r_{\rm i}$, 
with a poloidal extension from 0 to 0.5 radians,
and an azimuthal extension from 0 to 2$\pi$ radians. 
The second control volume, CV2, covers a radial range from $12\,r_{\rm i}$ to $22\,r_{\rm i}$,
with the same poloidal and azimuthal extension.
These control volumes are designed to facilitate the analysis of mass fluxes across different regions in the disk.

\begin{figure*}
\centering
\includegraphics[width=17cm]{\figurepath/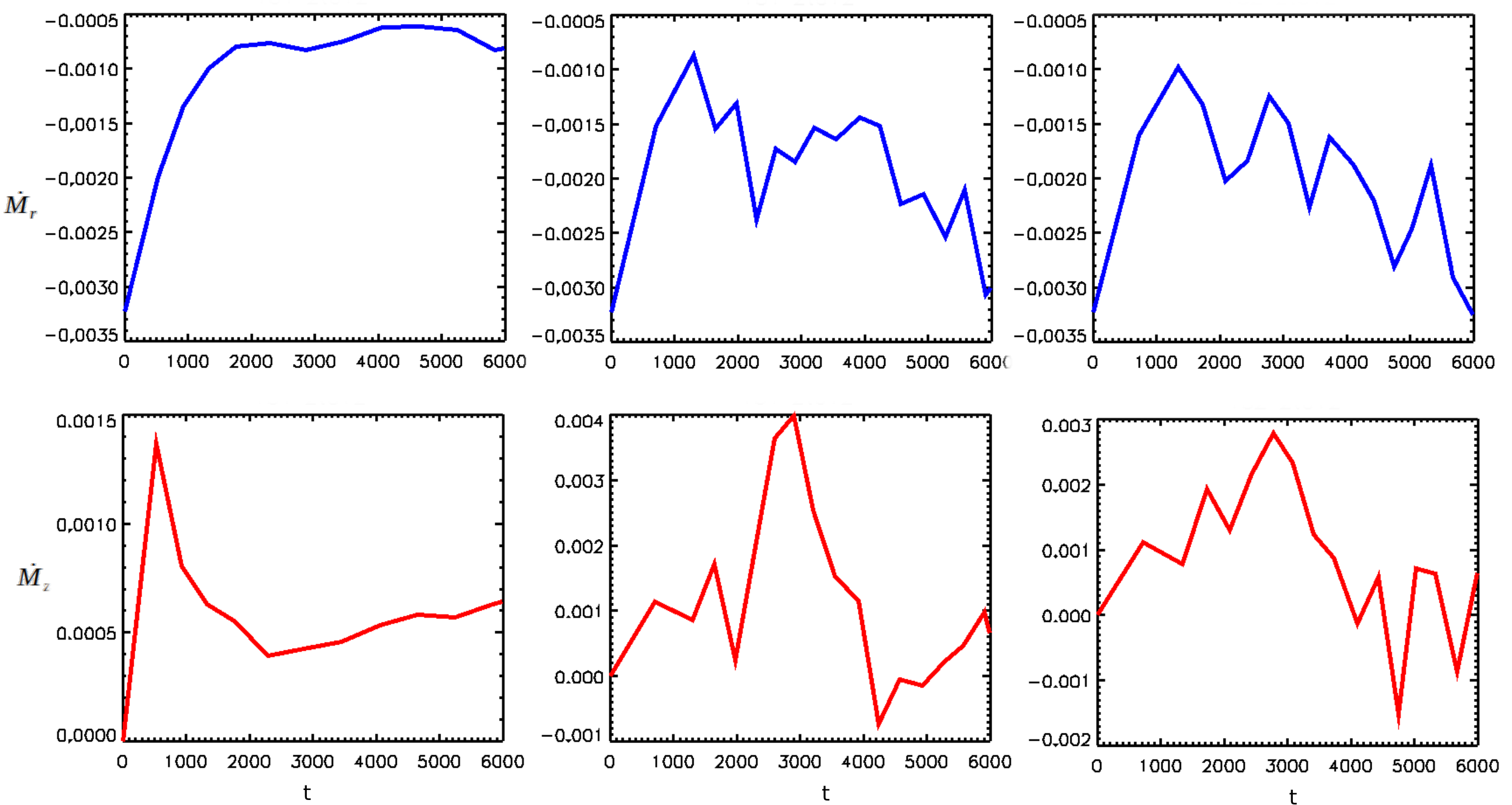}
\centering
\caption{Comparison of the mass flux evolution.
The panels show the evolution of the accretion $\dot{M}_r$ (top) and outflow $\dot{M}_z$ (bottom, upper hemisphere only) mass fluxes,
as measured from control volume CV1, 
for the simulations involving a single star (rs1) and for a binary system with mass ratio $q= 0.01$ (rb1) and $q= 0.2$ (rb2), respectively.
The accretion rates are expressed in units of $\dot{M}{\rm i}$.}
\label{fig:massflux_comparision}
\centering
\includegraphics[width=17cm]{\figurepath/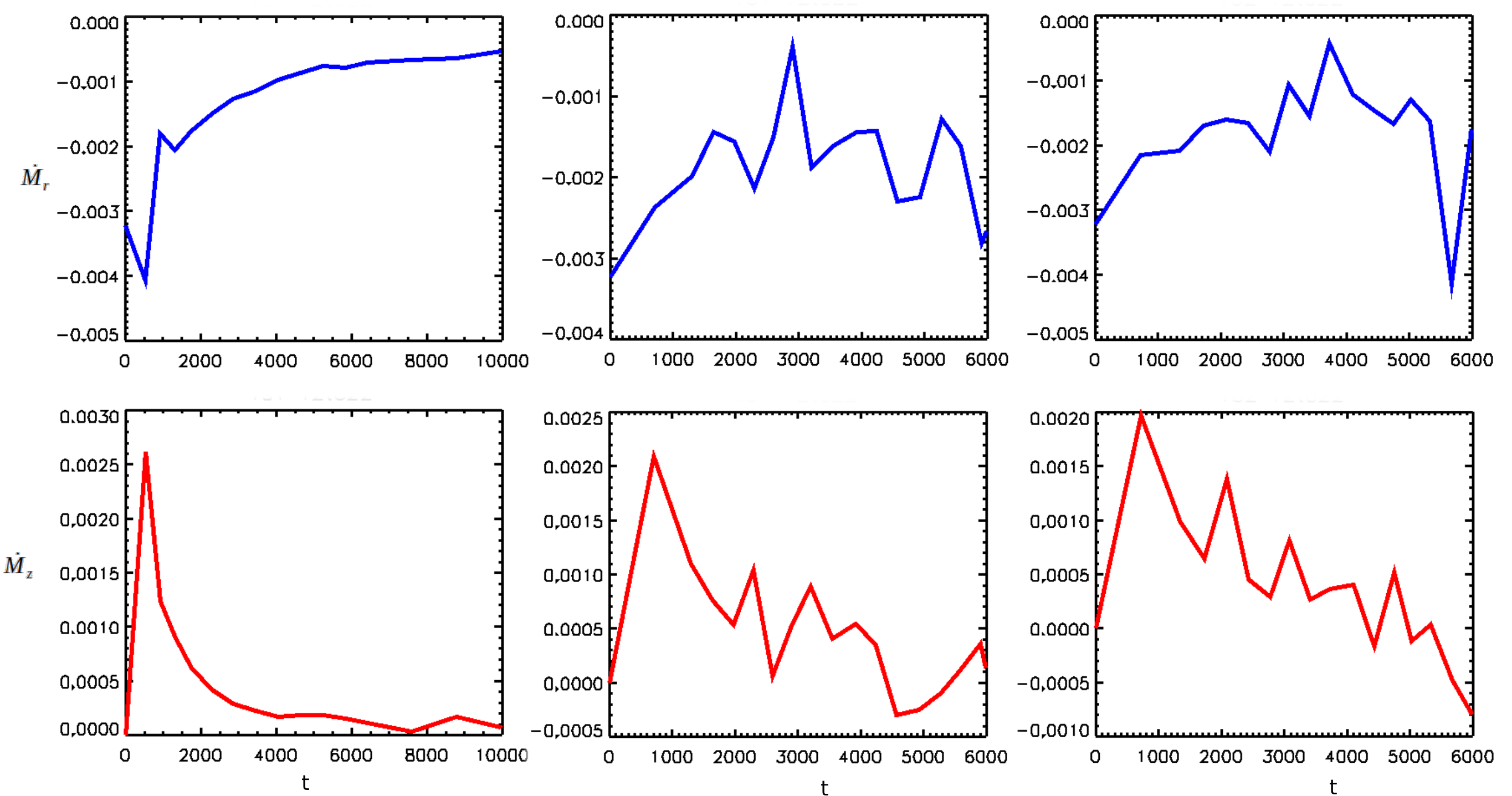}
\centering
\caption{Similar to Fig.~\ref{fig:massflux_comparision}, but measured for control volume CV2.}
\label{fig:massflux_10to20}
\end{figure*}

In Figure~\ref{fig:massflux_comparision} we display the evolution of accretion and outflow mass fluxes measured from a similar control volume
in simulations involving both a single star and a binary system with a mass ratio of 0.2.
In addition, Figure~\ref{fig:massflux_10to20} shows similar mass fluxes, however derived from the second control volume CV2.

In summary, we obtain the following results
(see Figures~\ref{fig:massflux_comparision} and ~\ref{fig:massflux_10to20}).
The accretion and outflow mass fluxes evolve smoothly in the simulation for the single star setup. 
For the binary star setup, these variables considerably fluctuate.
These fluctuations are caused by the tidal forces arising in the time-dependent Roche potential.
While these tidal effects generate features such as spiral arms in the disk, they also significantly disrupt the typical accretion pattern in the disk. 
Moreover, regions in the outflow and near the disk exhibit asymmetric structures like enhanced density (see Figure~\ref{fig:2Dview_rho_rb2_q0.2}) that arise from the time-dependent Roche potential.

Additionally, we observe that the accretion and ejection mass fluxes in some regions show some peaks, indicating that the accretion and outflow mass fluxes are enhanced for short time intervals.
This confirms the discussions in our recent paper \citep{2022ApJ...925..161S} that the angular removal from the disk 
and thus the tidal effects on the disk-jet are more efficient when a secondary star is present.
Again we see (as discussed in \citep{2022ApJ...925..161S}) that the asymmetric features present in the disk
are transferred to the outflow by the launching process.

Thus, essentially, the jet carries signatures of the disk evolution, and the primary effects, such as the spiral 
structures we observe in the outflow, are imprinted by the structural evolution of the disk.
However, we see also other features in the outflow near the launching site, which can be directly attributed to the asymmetric (and time-dependent) Roche potential, such as the formation of an arc-like structure in the outflow, and a perturbation in the velocity pattern \citep{2024ApJ...966...82S}.

\subsection{Outflow dynamical properties}
\label{outflow dynamic}
We now briefly discuss the essential properties of the disk wind concerning dust entrainment.
It is commonly accepted that protostellar disks are cool and therefore able to host dust grains out to a certain radius.

Compared to jets, disk winds are less extended and can thus be infrequently resolved.
While jets can be detected by imaging, disk winds are merely detected by spectral lines.
For instance, \citet{2023NatAs...7..905F} detected emission from disk winds, particularly in the low-velocity component of the 
[OI]~$\lambda$~6300 line. 
Spectral mapping indicated enhanced emission which allowed to fit extended wind emission applying toy modeling assuming
an inner radius of 0.08 au and a possible outer radius beyond 40 au.

For the same line, \citet{2023A&A...670A.126F} provided MUSE observations for a number of sources, 
giving evidence for line emission originating in the inner parts of a protoplanetary disk or inner disk winds.
Evidence for MHD disk winds was further provided by spectroastrometry of optical forbideen emission lines by \citet{2021ApJ...913...43W}.
\citet{2024AJ....167..127B} reported the first spatially resolved [Ne II] disk wind emission from T Cha, 
revealing an extended structure beyond the disk continuum.

Extended jets from young stars also produce forbidden emission lines, excited in internal or external shocks.
While this radiation can be mapped by imaging, this approach is, so far, not easily possible for the launching site for 
either disk winds or jets on scales below say several tens of AU.

However, observations of jets at larger distances allow for the estimation of essential jet parameters.
For example, the jet densities can be determined from shock-excited forbidden emission lines in DG\,Tau 
\citep{1993ApJ...410L..31S} as well as temperatures in combination with dynamical modeling \citep{2003A&A...410..155P}. 

From the theoretical side, MHD jets and winds are considered as {"}cold{"} flows.
Theoretical studies \citep{1982MNRAS.199..883B, 1983ApJ...274..677P} and numerical simulations of jet {\em formation} 
typically neglect the thermal pressure for the wind motion 
\citep{1985PASJ...37..515U, 1995ApJ...439L..39U, 1997ApJ...482..712O, 1999ApJ...526..631K, 2006ApJ...651..272F}.
In agreement with the jet formation studies, theory, and simulations of jet {\em launching} - investigating the disk-jet transition - confirm these results.
The cool disk material is lifted out of the disk, becomes accelerated, and, while further expanding, does further cool
down\footnote{Typically a polytropic gas law is considered.}
\citep{2002ApJ...581..988C,2007A&A...469..811Z,2012ApJ...757...65S}.

The initial (disk surface) gas temperature, (respective pressure) is an important parameter for the wind motion, 
as it determines the mass flux of the wind or jet.
This is already a consequence of steady-state MHD theory \citep{2007prpl.conf..277P}.
MHD simulations of jet formation including radiation pressure have shown that the disk material can become
heated when rising from the disk - resulting in a larger mass flux, which moves with a lower speed \citep{2011ApJ...742...56V}. 
Still, the outflow remains {"}cool{"} \citep{2003astro.ph.11621F}.
 
Only when gas is shocked downstream and heated to approximately 3000-4000 K, it becomes visible as jet knots.  
Hydrodynamic Simulations by \citet{2001ApJ...557..429L} find shock temperatures up to $10^4$\,K while applying 
jet temperatures of 270K (following \citet{1999A&A...343..571G}; see also \citet{Rabenanahary2022} for an 
overview over typical jet parameters).

Various studies explore dust continuum observations within the disk wind and different regions of the disk.
For example, \cite{2024A&A...686A.201N} presented high angular resolution ($\simeq 30 AU$) ALMA observations of 
the emblematic L1448-mm protostellar system and found suggestive evidence for an MHD disk wind.
The disk seen in the dust continuum (0.9 mm) has a radius of $\simeq 23$ AU, and they measure the maximum brightness temperature of the continuum 
about 110 K at this region.

In addition, \cite{2022A&A...668A..25V} studied the accreting intermediate-mass T-Tauri star RY Tau, which
exhibits an active jet and a known disk wind. They discussed how archival optical data and recent near-infrared observations 
of the RY Tau system reveal two horn-like components extending as a cone from RY Tau. 
While scattered light from the disk surrounding RY Tau is observable in the near-infrared spectrum, it remains unseen 
in optical wavelengths, suggesting the potential presence of scattered light originating from a disk wind. 
By applying radiative transfer modeling, they found that small grains that are elevated above the disk surface can reproduce 
the observed effects.

\cite{2023A&A...678A...6L} studied the circumstellar material around SU Aur, focusing on disk geometry, composition, and inner
dust rim structure and confirmed the presence of a dusty disk wind, strengthened by a recent infall event, 
which also causes significant misalignment between the inner and outer disks.

Other studies demonstrated that the dust grains carried away by the gas component of the wind survive in contact with 
the hot $(10^4 K)$ gas \citep{2008AstL...34..231T}.

\citet{2022A&A...659A..42R} have further discussed and approved the idea that despite heating of the dust grains that are entrained in 
the high-temperature gas flow these dust particles do not reach the dust sublimation threshold and would thus remain solid along their trajectories.

\begin{figure}
\centering
\includegraphics[width=0.9\columnwidth]{\figurepath/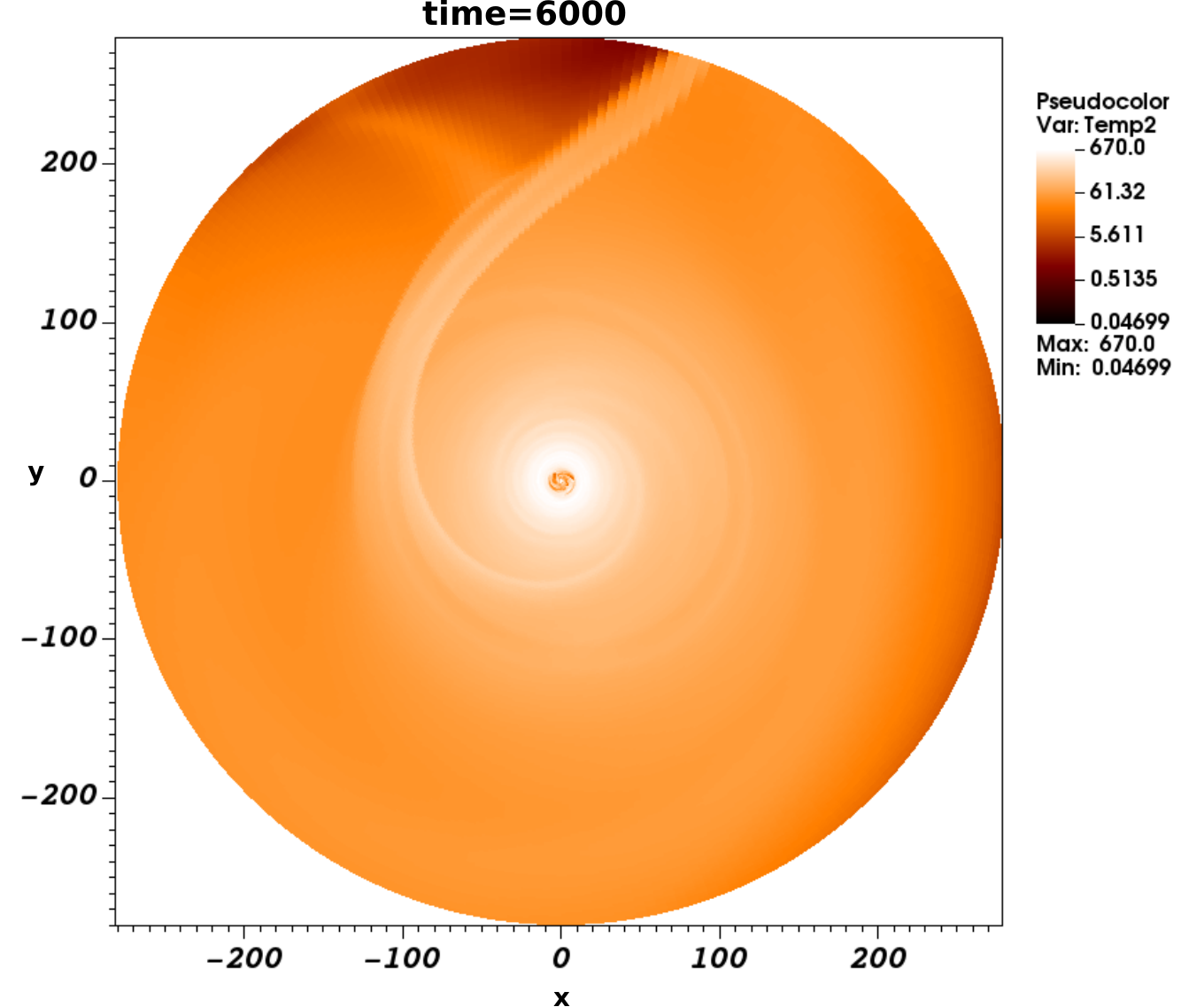}
	\caption{Temperature distribution of a circumstellar disk in a binary system.
    The map displays a snapshot of the gas temperature (in Kelvin) in the midplane of a circumstellar disk
    in a binary system (mass ratio of \( q = 0.2 \)) at $t= 6000$.}
	\label{fig:Temp_bin_plot_t6000}
\end{figure}

\begin{figure*}
\centering
\includegraphics[width=18cm]{\figurepath/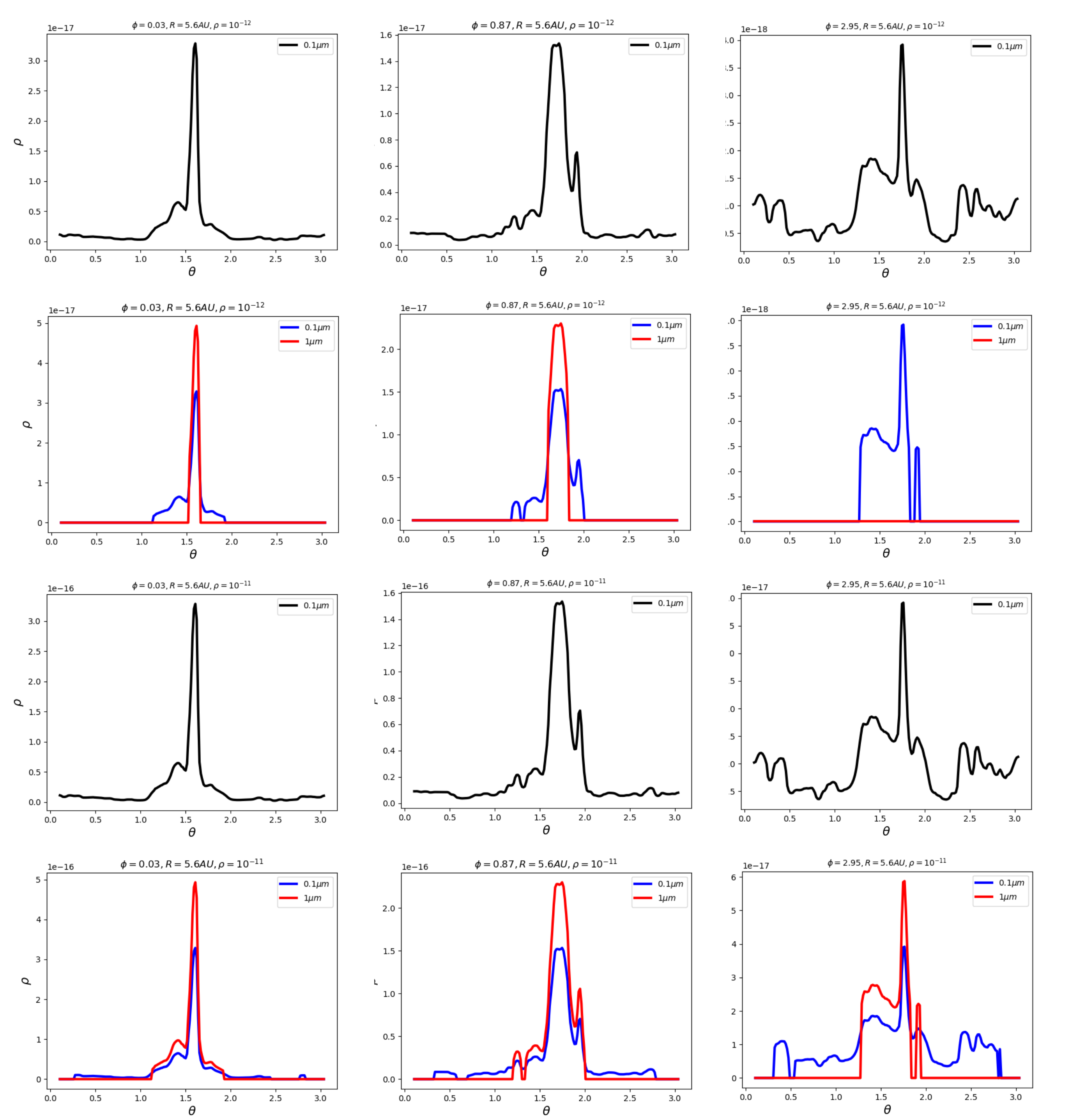}
\caption{profiles of the dust density as a function of the polar angle $\theta$.
Shown are profiles for silicate dust particles (0.1 $\mu$m and 1 $\mu$m) at 5 AU and for varying azimuthal angles $\phi$, illustrating our two approaches.
The top row represents the {\em SDD} approach, considering only 0.1 $\mu$m particles, well-coupled to the gas with a dust-to-gas ratio of 0.01.
The second row shows  the {\em SND} approach, incorporating both 0.1 $\mu$m and 1 $\mu$m particles, distributed according to their Stokes number in the disk-jet system, applying the gas density $\rho = 10^{-12}\, gr \, cm^{-3}$.
The third and the last rows present the {\em SDD} and {\em SND} approach, respectively, for a higher gas density of $\rho = 10^{-11}\, \rm gr \,{cm}^{-3}$.}
\label{dust_prof_twosizes_5Au}
\end{figure*}

\section{Radiation transfer modeling}
\label{Radiation modeling}
One of the aims of this paper is to generate a dust continuum radiation map of the accretion-ejection structure using a radiative transfer code.
Given that we have a comprehensive three-dimensional simulation of the jet formation from a circumstellar disk in both single and binary star systems, 
the RADMC3D code \citep{2012ascl.soft02015D} provides the tools to derive the desired dust continuum radiation map corresponding to our results
from the dynamical simulation. 
In this section, we describe different steps that we perform to have a proper and reasonable radiation map.

It is important to note that our objective is not to directly compare our findings with a specific observational object, but rather provide insight into how the structures identified in our simulation may manifest in observations. 
Additionally, we aim to suggest potential instruments that could be utilized to figure out the presented features such as the disk wind or spiral arms.

As a first step, we examined the gas temperature in our simulation to determine if dust particles could potentially exist. 
To estimate the gas temperature, we apply again a polytropic gas law,
\begin{equation}
T = T_0 \rho^{\gamma-1}
\end{equation}
where $T_0 \equiv p_0 \mu m_{\rm H}/\rho_0\,{\rm K}$. 
Here, $\mu$ is the average mass per particle and $m_{\rm H}$ is the mass of a hydrogen atom and $k_{\rm B}$ is the Boltzmann constant.
With this, we can now define the fiducial gas temperature in code units and then further convert it to astrophysical units, 
e.g. for a young stellar object \citep{2013MNRAS.428.3151T},
\begin{equation}
T|_{z=0} = 10^4 \left(\frac{\epsilon}{0.1}\right)^2 \left(\frac{M}{M_\odot}\right) \left(\frac{r}{0.1 \text{ AU}}\right)^{-1}.
\label{tem_physical_unit}
\end{equation}
where the parameter $\epsilon$ denotes the initial thermal height scale.
In Figure~\ref{fig:Temp_bin_plot_t6000}, we present a snapshot of the gas temperature in the disk midplane for the simulation of a binary star system (with a mass ratio of $q=0.2$) at time $t=6000$. The map indicates that the disk temperature remains below the sublimation temperature of silicates, which is approximately 1200 \,K.
This is supported by \cite{2017A&A...604A...5H} who investigated the effect of the grain radius distribution on dust temperature in circumstellar disks, finding values within a range similar to those obtained in our study. Therefore, we conclude that the conditions are suitable for the survival of silicate dust particles.

It is important to note that we do not incorporate the energy equation in our simulation setup. 
Instead, we derive the temperature simply using a power law relation in the polytropic equation of state.
On the other hand, it is well known that jet material originates from the cold, low-entropy accretion material in the
disk \citep{2002ApJ...581..988C,2007A&A...469..811Z,2012ApJ...757...65S}.

Consequently, the resulting outflow exhibits low temperature, enabling the survival of small dust particles in certain regions of the wind.

Some of the recent studies have investigated the circumstances under which dust particles can persist in the wind
\citep{2008AstL...34..231T,10.1093/mnras/stab090,2022A&A...659A..42R}.
For example, \cite{2008AstL...34..231T} examined the survival of dust grains in the disk wind of T Tauri stars (TTSs) and the conditions under
which dust can survive even in the hot wind component.
Furthermore,\cite{2022A&A...659A..42R} analyzed the behavior of dust grains in cold magnetic winds and photo-evaporated warm ionized winds, 
discovering that dust grains in the cold magnetic wind tend to follow a shallower trajectory compared to the warm, ionized winds.

In the following, we explain our approach for generating radiation maps of a series of disk-jet structures, specifically focusing on the radiation pattern that is emitted by the dust in the outer part of the disk.

We incorporated the radiative transfer code RADMC3D \citep{2012ascl.soft02015D}\footnote{ See \texttt{https://github.com/dullemond/radmc3d-2.0}}.
RADMC3D code performs a thermal Monte Carlo simulation to compute the dust temperature under the assumption that the dust is in radiative 
equilibrium with its radiation field.
To create a radiation map of the dust continuum, we need information on the distribution of dust particles in both disk and jet.
Because our MHD simulation does not account for dust, we require to model dust using the gas density.

Therefore, we employ two different dust density approaches. In the first approach, we consider only small dust particles of size 0.1~$\mu$m, which are well-coupled to the gas and follow a dust-to-gas density ratio of 0.01 (hereafter referred to as Smooth Dust Density -  {\em SDD} approach).
To calculate the dust density, we simply divide the gas density by 100. In the second approach, we assume the dust has had sufficient time to settle and grow. 
We consider two dust sizes with 0.1~$\mu$m and 1~$\mu$m, distributing them based on their Stokes number
(hereafter referred to as {\em Stokes Number for Dust Density} - {\em SND} approach). The Stokes number characterizes the tendency of the dust to flow with the fluid. Inspired by the results of \cite{2022A&A...659A..42R}, we distribute the dust only where the particle Stokes number is up to 0.1. 
Figure~9 of that study shows that the normalized dust density drops by one order of magnitude for Stokes number below 0.1. To take into account the settling of larger dust grains, we multiply the gas density by 0.015 for the larger dust grains.
\begin{figure*}
\centering
\includegraphics[width=14.5cm]{\figurepath/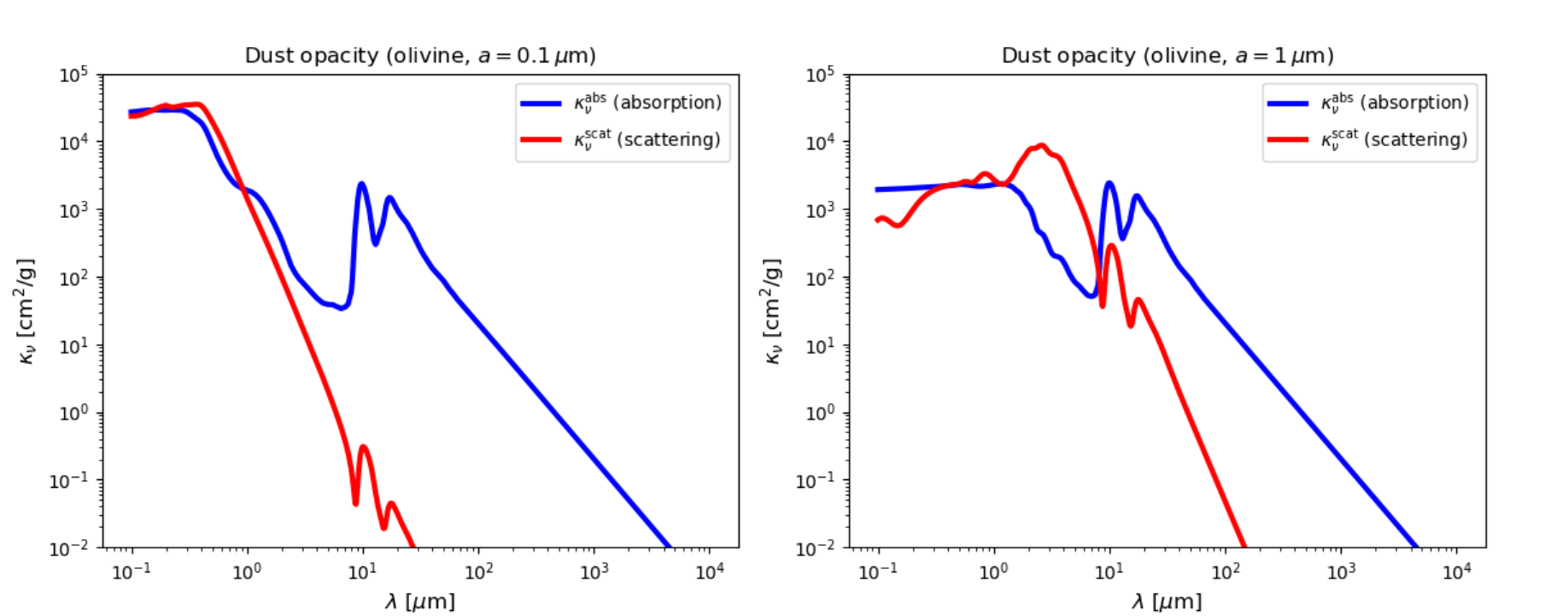}
\caption{Dust opacities. 
Shown is the dust opacity as a function of wavelength for silicate particles with sizes 
of 0.1 $\mu$m (left) and 1 $\mu$m (right), respectively. 
The values displayed represent the absorption coefficient per unit mass (blue) and the 
scattering coefficient per unit mass (red) along wavelength.}
\label{opacity_0.1micron}
\end{figure*}

To clarify our recipe for the second approach, we first calculate the Stokes number for each cell as
\begin{equation} \label{St}
{\rm St}=\Omega_{\rm K} t_{\rm stop} = \Omega_{\rm K} \frac{\rho_{\rm d} s_{\rm d}}{\rho_{\rm g} v_{\rm th}},
\end{equation}
where $\Omega_{\rm K} = \sqrt{GM_{\star}/(R\sin{\theta})^3}$ is the Keplerian angular velocity with $M_{\star}$ being the stellar mass. 
The $t_{\rm stop}$ is the stopping time of the dust particle (the time for the exponential decay of the particle
relative velocity due to the gas drag) with size $s_{\rm d}$. 
In Eq.~\ref{St}, $v_{\rm th}=\sqrt{8 k_{\rm B} T_{\rm g}/\mu m_{\rm H} \pi}$ is the gas thermal velocity.
We use $\mu=2.3$, $M_{\star}=1M_{\odot}$ and Eq.~\ref{tem_physical_unit} for calculating the gas temperature.
Then, we distribute the dust as
\begin{equation}
\rho_{\rm d,s} = 
\begin{cases} 
      0 & {\rm St}>0.1 \\
      f_{\rm dg} \rho_{\rm g} & {\rm otherwise}
   \end{cases}
\end{equation}
where we set $f_{\rm dg}$ to 0.01 for $s=0.1\mu$m and 0.015 for $s=1\mu$m.
 
Figure~\ref{dust_prof_twosizes_5Au} presents the dust density profile along the polar angle $\theta$ for the distribution of silicate dust particles at $R=5.6$~AU and two different values for the reference gas density. 

The top row presents the  {\em SDD} approach, considering only 0.1 $\mu$m particles, well-coupled to the gas with a dust-to-gas ratio of 0.01. 
The second row shows the {\em SND} approach, incorporating both 0.1 $\mu$m
and 1 $\mu$m particles, distributed according to their Stokes number in the disk-jet system for the gas density of $\rho_{\rm i} = 10^{-12}\rm gr \,{cm}^{-3}$. 
The third and the last rows present the {\em SDD} and {\em SND} approach, respectively, for a higher gas density of $\rho_{\rm i} = 10^{-11}\, \rm gr \,{cm}^{-3}$. Each column belongs to a different azimuth.

The plot indicates that the larger dust (1 $\mu m$) tends to settle closer to the disk area, whereas the smaller dust reaches higher altitudes and are also locates inside the disk wind.
For a larger gas density (bottom row), the 0.1 $\mu$m dust particles reach even higher altitudes and become more widely distributed throughout the disk wind. We decided to ignore the millimeter dust particles because in our {\em SND} approach, they are well located in the midplane. Besides, in the case of $\rho_{\rm i} = 10^{-12} \rm gr \,{cm}^{-3}$, their Stokes number is above our threshold. Therefore, we expect that these large grain contribute only in the thermal radiation in long wavelengths that would be similar to the gas in the midplane. On the other hand, they have lower (and gray) opacity compared to smaller grains in smaller wavelengths that results in negligible absorption by these large particles.
Although considering a distribution of all possible sizes that are suggested by observation or a dust settling model would be more realistic, it is beyond the main scope of this study.
\begin{figure*}
\centering
 \includegraphics[width=16cm]{\figurepath/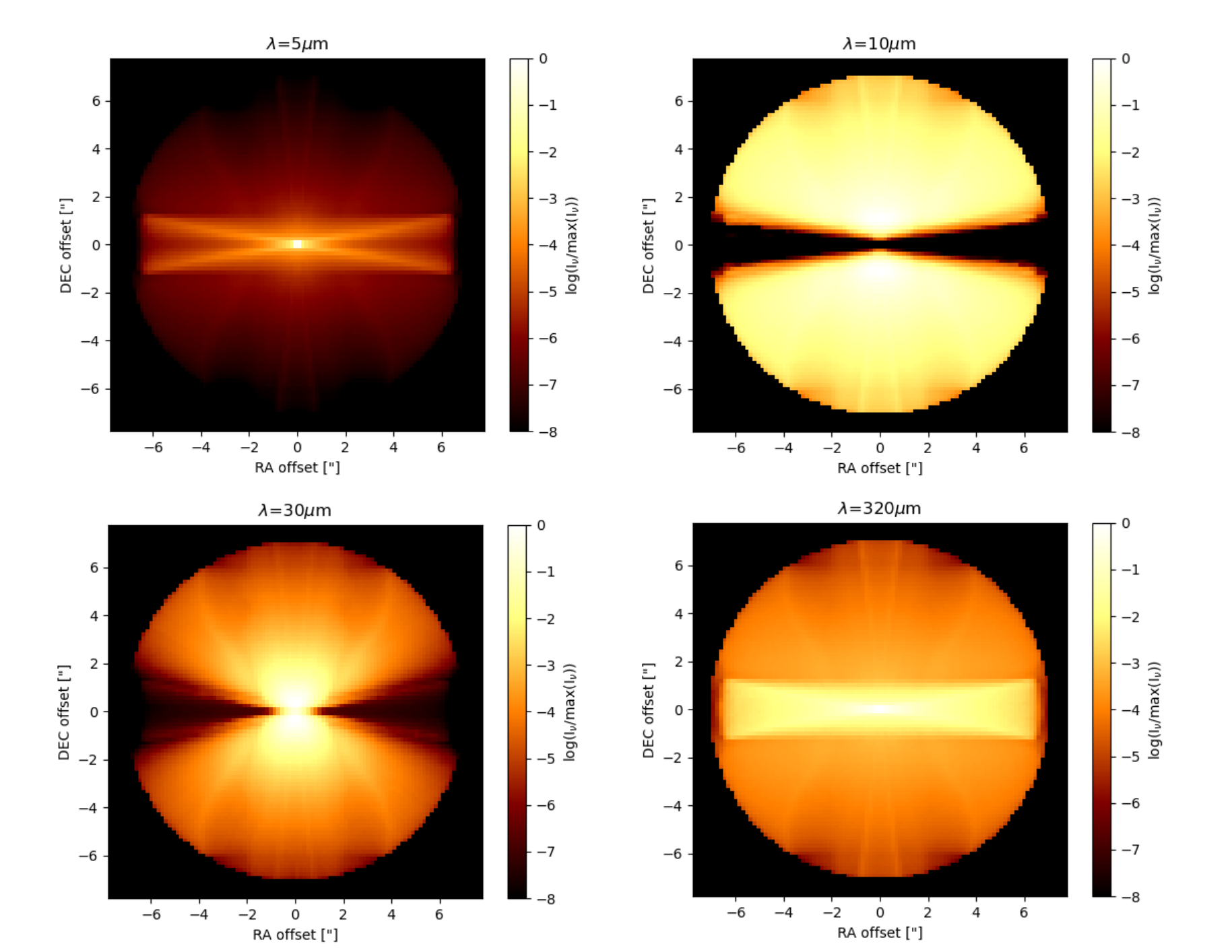}
\caption{Radiation map for a single star-disk-jet system.
 Shown are the maps of the dust continuum radiation at different wavelengths for the reference run considering a single star with disk and jet at dynamical time t=6000. 
 The physical gas density at the inner disk radius is chosen as $\simeq 10^{-12} {\rm g\,cm}^{-3}$.
 The snapshots correspond to a l.o.s. inclination angle $\delta=90$, and have been calculated for wavelengths 5, 10, 30, and 320 $\mu m$, assuming a distance of 140\,pc.
 The color bar indicates intensities in Jy/beam and is normalized based on the maximum and minimum intensity at the specific wavelength.}
\label{fig:runsin_large_simple3}
\end{figure*}
\begin{figure*}
 \includegraphics[width=18cm]{\figurepath/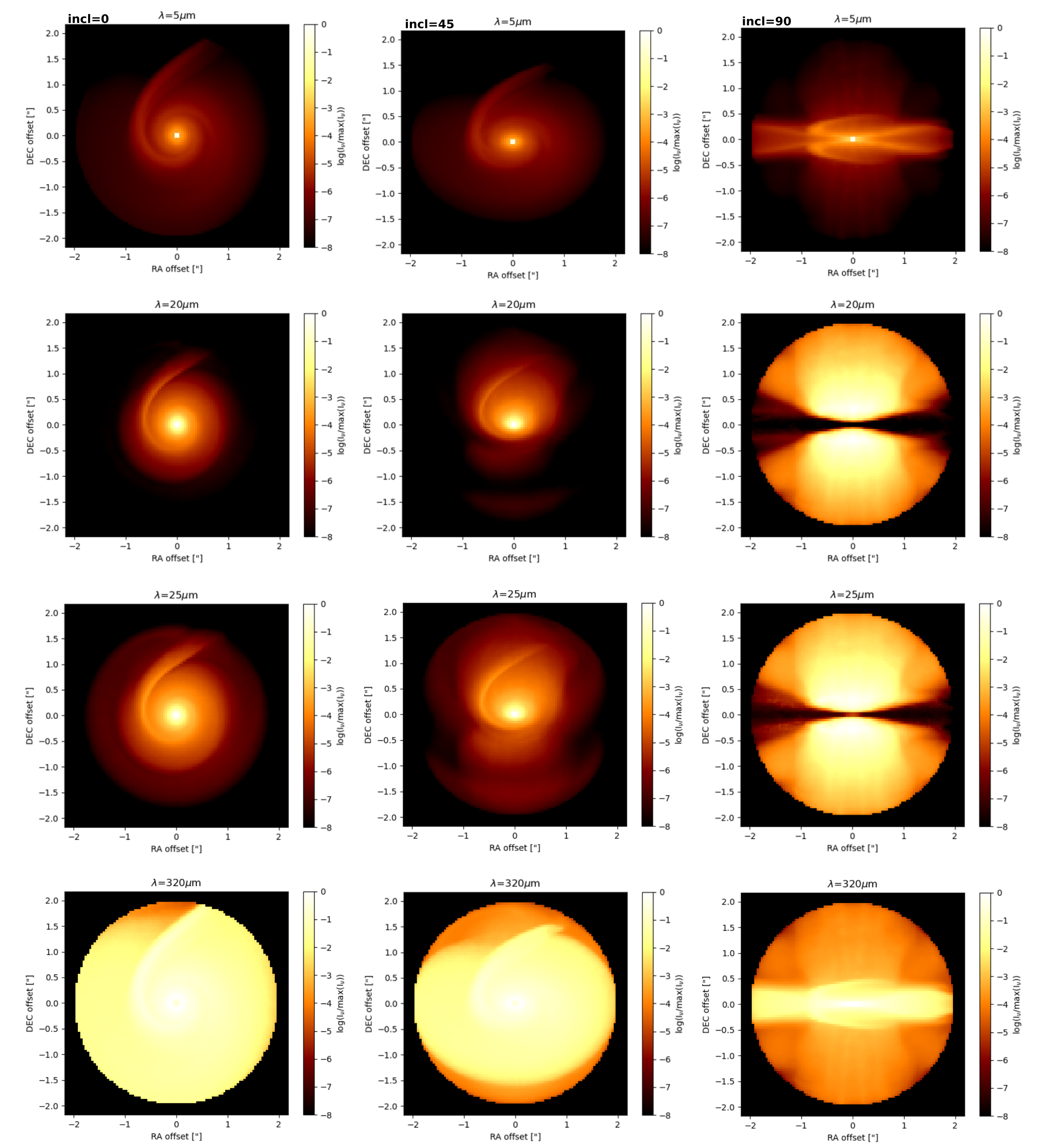}
\caption{Radiation map of the binary system. 
The maps of the dust continuum radiation at different wavelengths are shown for a simulation considering a 
binary star-disk-jet with a mass ratio of $q=0.2$ at dynamical time t=6000. These maps show various inclination angles relative to the observer's l.o.s. of $0, 45$ and $90$ degrees (from left to right).
A distance of 140\,pc is assumed. 
Here, the dust density is assumed to be $1\%$ of the gas density ({\em SDD} approach) and the physical gas density at the inner disk radius is chosen as $\simeq 10^{-12} {\rm g\,cm}^{-3}$. The color bar represents intensities in Jy/beam and is normalized based on the maximum and minimum intensity at the specific wavelength.}
\label{fig:rb2_simple_all_inc}
\end{figure*}
\begin{figure*}
\includegraphics[width=18cm]{\figurepath/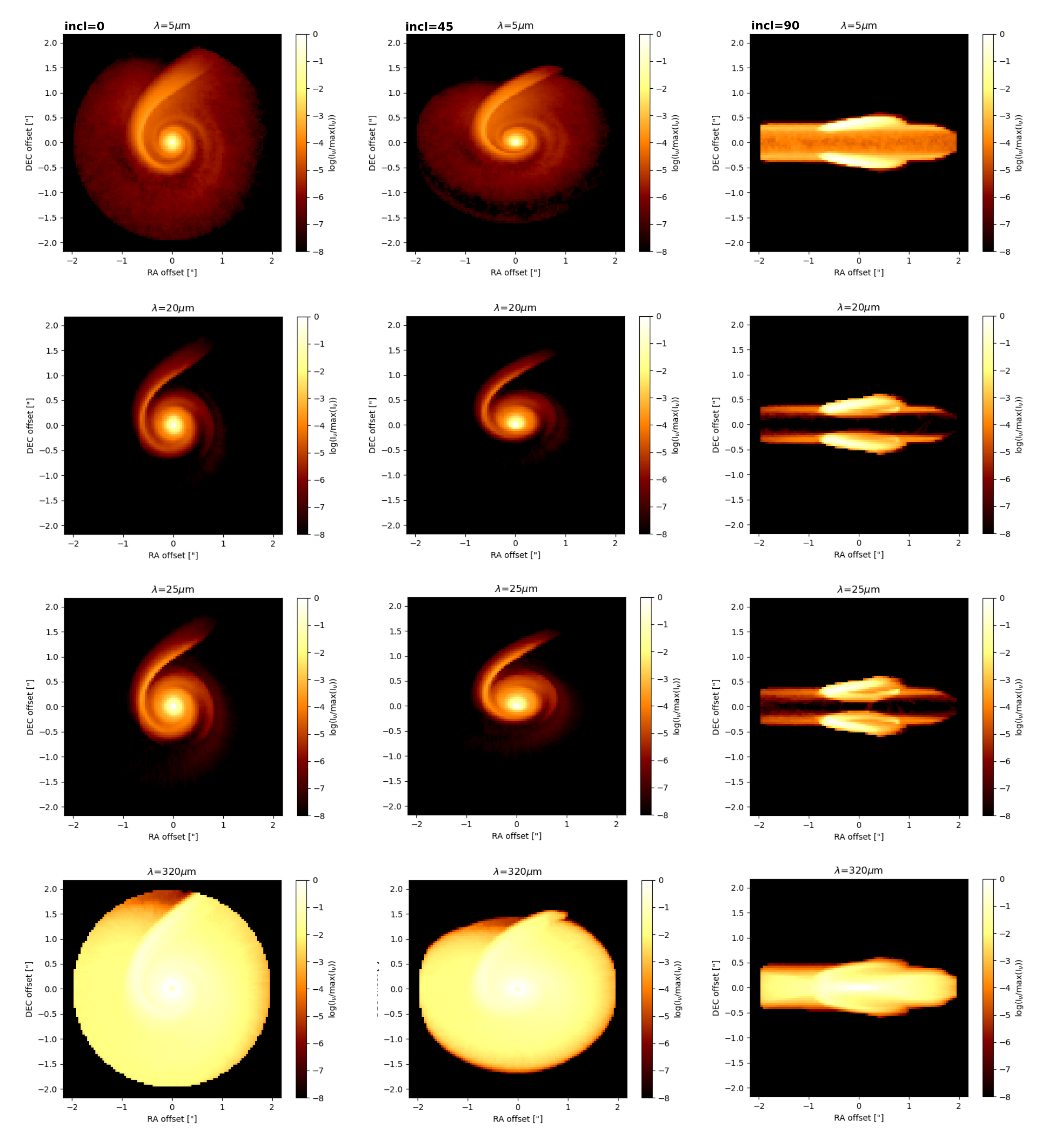}
\caption{ Radiation maps of the binary system. 
Maps of the dust continuum radiation at different wavelengths are shown for a simulation including a binary star-disk-jet 
with a mass ratio of q=0.2 at dynamical time t=6000. 
Here, the density profile is obtained according to the Stokes number of two dust particle sizes, 
specifically $10^{-5}$ and $10^{-4}$ cm.
The physical gas density at the inner disk radius is chosen as $\simeq 10^{-12} {\rm g\,cm}^{-3}$.
These maps show various inclination angles relative to the observer's l.o.s. of $0, 45$, and $90$ degrees (from left to right). A distance of 140\,pc is assumed.
The color bar represents intensities in Jy/beam and is normalized based on the maximum and minimum intensity at the specific wavelength.}
\label{fig:rb2_profile_all_inc}
\end{figure*}

Beside the dust spatial distribution, the opacity of the dust should be provided for the code. 
The opacity profile across the entire system plays a vital role in computing the radiation emitted by the dust particles.
The opacity is directly related to the intensity, which is defined by 
\begin{equation}
 I=I_0 e^{-\kappa \rho d}=I_0 e^{-\tau}
 \label{optical_depth}
\end{equation}
where $\tau$ is the optical depth, $\kappa$ is the opacity or the absorption coefficient, $I_0$ is the
incident intensity, $\rho$ is the density and $d$ is the distance that a photon travels in the gas. The opacity of a medium containing dust particles is determined by factors such as the size and composition of
the dust particles, as well as the wavelength of the photon. 

To provide the opacity for the code, we consider the dust as Olivine grains with $\rho_{\rm s}=3.71{\rm gr/cm^3}$. The dust opacities for each size are calculated using the data from Jena database~\footnote{https://www.astro.uni-jena.de/Laboratory/OCDB} \citep{1994A&A...292..641J,1995A&A...300..503D}
and are specified in a separate file for each size.

In Figure~\ref{opacity_0.1micron}, we illustrate the opacity profiles across different wavelengths
$\kappa_\nu(\lambda)$, derived for an Olivine dust particle with a size of 0.1 $\mu$m (left) and 1  $\mu$m (right). 
As shown in Figure~\ref{opacity_0.1micron}, the absorption decreases with increasing wavelength, except for
a peak at 10 $\mu$m.
Therefore, at 10 microns, we expect a weaker emissivity, as significant absorption occurs at this wavelength for silicate dust particles.

We used $10^{7}$ incident photons and about $10^{7}-10^{8}$ scattered photons, depending on the wavelenght, for the models with $\rho_{\rm i} = 10^{-12}\, \rm gr \,{cm}^{-3}$. However, we increased the number of scattered photons to $10^{8}$, and even to $10^{9}$ at the wavelength of 5 $\mu$m for $\rho_{\rm i} = 10^{-11}\, \rm gr \,{cm}^{-3}$ in order to obtain smooth and less spiky radiation maps.

We note that in the present simulations, we do not include the dust evolution in the
model setup. 
Instead, a predefined distribution for dust particles is assumed and considered to 
be in thermal equilibrium with the gas. 
Indeed, a promising direction for future work is to incorporate dust evolution 
into the modeling, 
potentially enabling a more realistic and self-consistent approach when
simultaneously evolving dust and gas.

\section{Radiation maps} 
Dust continuum radiation maps of disk-jet systems may provide insights into their structural composition. 
These maps capture the emission from dust grains present within the disk wind and the underlying disk, offering essential information on the dust 
distribution, temperature profile, and overall system outflow structure. 
Additional information on gas dynamics and velocity may be obtained, although gas line radiation is required for that type of analysis.

\subsection{Circumstellar disk-jet in a single star system}
In this section, we present the radiation map of the disk-jet structure in a single star system that has achieved a steady state.
To generate the images, we employed the {\em SDD} approach, considering small dust particles of sizes of  0.1~$\mu$m 
which are well-coupled to the gas and follow the dust-to-gas density ratio of 0.01.

Figure~\ref{fig:runsin_large_simple3} shows the dust continuum radiation maps from our
simulations, showcasing the circumstellar disk that initiates the jet in a single star 
at a dynamical time of t=6000.
These snapshots show the radiation map of the disk-jet system at inclinations of 90 degrees, relative to the line of sight, 
at wavelengths, 5, 10, 30, and 320 microns.
The images represent the intensity in units of Jansky per beam $(Jy/{\rm beam})$ assuming
a source distance of 140\,pc.

In order to better understand and interpret our radiation maps, we may consider the silicate 
opacity as a function of wavelength, as illustrated in Fig.~\ref{opacity_0.1micron}. 
The opacity values displayed represent the absorption coefficient per unit mass (in blue) and the scattering coefficient per unit mass across the wavelengths.
The opacity profile shows a peak around 10 $\mu m$, which is known as an important 
spectral feature of silicates, and then slowly levels down for larger wavelengths. 

Therefore, the disk midplane, which has a larger dust density compared to higher $z$, 
is opaque for $\lambda=10\mu$m , but is almost transparent for wavelengths around $5\mu$m. 
The cross shape seen in $5\mu$m is indeed caused by scattering from the disk surface. 
This scattered radiation could leave the disk due to the reduced absorption and 
thus reach the observer.

The wind from the surface can be seen more easily in the $30\mu$m.
In longer wavelengths, for example, $320\mu$m, the disk is brighter because of the dust thermal emission.
\subsection{Circumstellar disk-jet in a binary star system}
In this section, we present radiation maps for a circumstellar disk-jet in a binary system.
To generate the radiation map of the disk-jet in a binary system, we employ both approaches outlined in section~\ref{Radiation modeling} and then compare the differences.

First, we consider the {\em SDD} approach, which accounts only for small dust particles with a size of 0.1~$\mu$m into account. 
These particles are well-coupled to the gas and follow the dust-to-gas density ratio of 0.01. 
Figure~\ref{fig:rb2_simple_all_inc} displays the dust continuum radiation for this model for various inclination angles.

The 5 $\mu m$ radiation map distinctly shows the scattered light from both the disk surface and the outflow. 
The spiral structure on the upper layer of the disk is evident for all inclination values. 
In the edge-on view, the spirals can be seen on both the top and the bottom sides. 
The outflow and the disk asymmetry can be also distinguished.

In the radiation maps for $20$ and $25\mu$m we see the scattering off the stellar radiation from the spiral on the 
layers closer to the surface than in the $5\mu$m, while the disk mid-plane remains completely opaque. 
However, because there is a drop in the opacity of Olivine around $25 \mu$m, photons can penetrate deeper into the 
disk and its outer part are visible in this wavelength. 

Spirals are also recognizable in $320\mu$m which shows the thermal emission of the disk mid-plane. 
There is also some radiation in this wavelength from the dust in the outflow.

While the spiral arms are a prominent feature in the disk mid-plane, these characteristics also manifest themselves in the 
upper layers of the disk and are also transported into the outflow, thus offering a robust validation tool for forming spiral
arms in the disk wind as well.
As the wavelength of the stellar radiation increases, photons can penetrate the deeper layers of the disk.

Figure~\ref{fig:rb2_profile_all_inc} is similar to Fig.~\ref{fig:rb2_simple_all_inc} but is based on the {\em SND} approach.
In comparison to the single-size model for dust particles, we find that most dust particles remain in the disk, resulting in a lower 
concentration of dust in the disk wind. 
Consequently, we receive radiation primarily from the lower part of the disk wind, while the majority of the radiation originates from 
the disk itself. 
In fact, we hypothesize that wind would be better observable in disks where the 
dust is less settled and not yet grown.

Considering Figure~\ref{fig:rb2_profile_all_inc}, we recognize that the larger dust particles remain in the disk.
The spiral arms are visible at 5, 20, and 25 $\mu$m, as well as at longer wavelengths.
In the {\em SND} approach, even small dust grains primarily reside in the disk, and partly lower layers of the disk wind, as their large Stokes number results from the low gas density in the wind.

Thus, in summary, the wind becomes observable when small dust is coupled to the gas, while the spiral arms are detectable when the dust has settled.
\begin{figure*}
\centering
 \includegraphics[width=17cm]{\figurepath/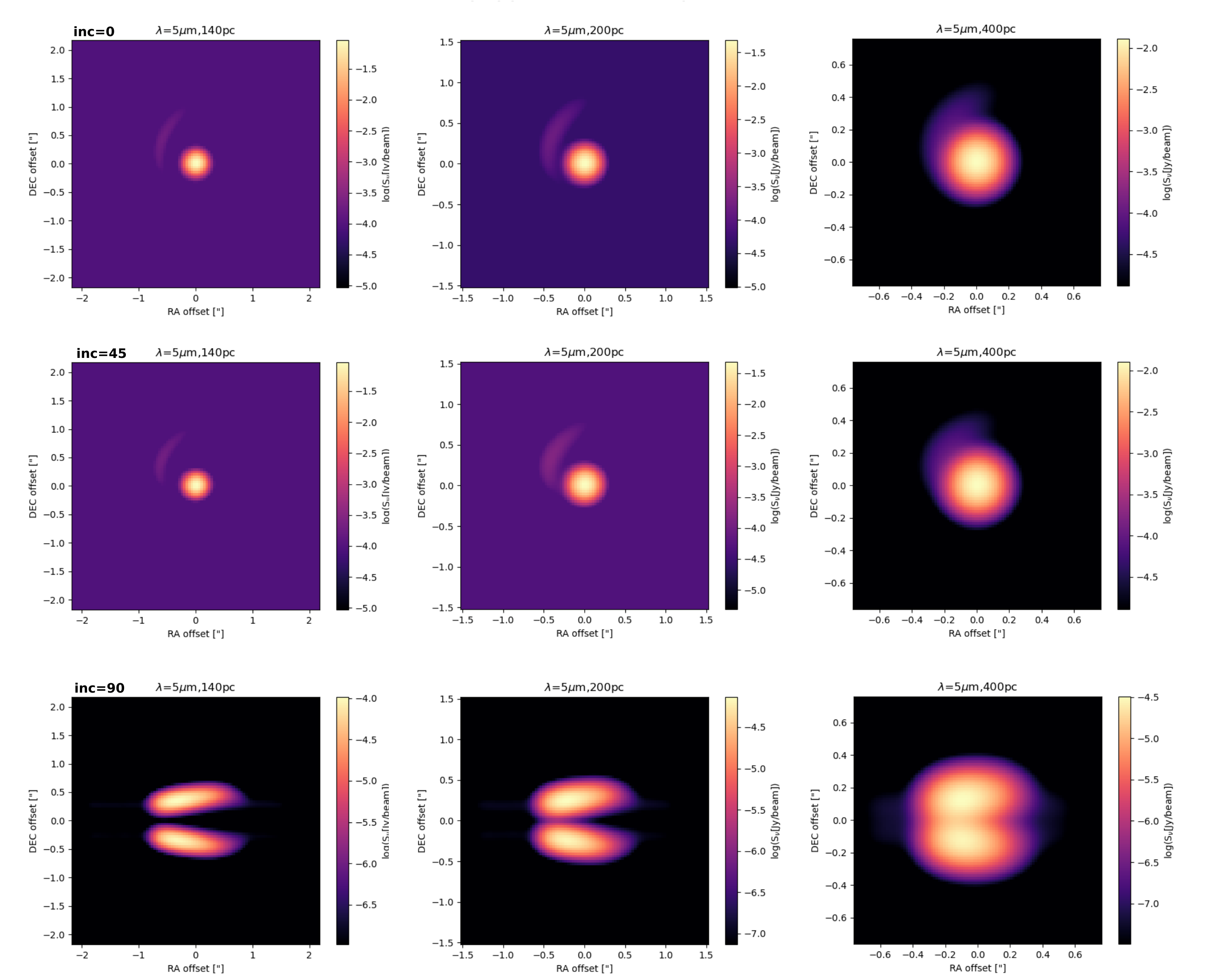}
\caption{Convolved radiation map for 5\,$\mu m$. 
The image displays the results of applying a convolution by assuming a PSF and using the MIRI 
instrument on the JWST and for distances of 140, 200, and 400 pc, respectively, and for l.o.s. inclination angles of 0, 45, and 90 degrees.
Here, the physical gas density at the inner disk radius is chosen as $\simeq 10^{-12} {\rm g\,cm}^{-3}$.
The color bar represents intensity in Jy/beam and is normalized based on the maximum and minimum intensity 
at the specific wavelength.}
\vspace{1cm}
\label{fig:rb2_convolved_MIRI}
\end{figure*}

\begin{figure*}
\centering
\includegraphics[width=14.4cm]{\figurepath/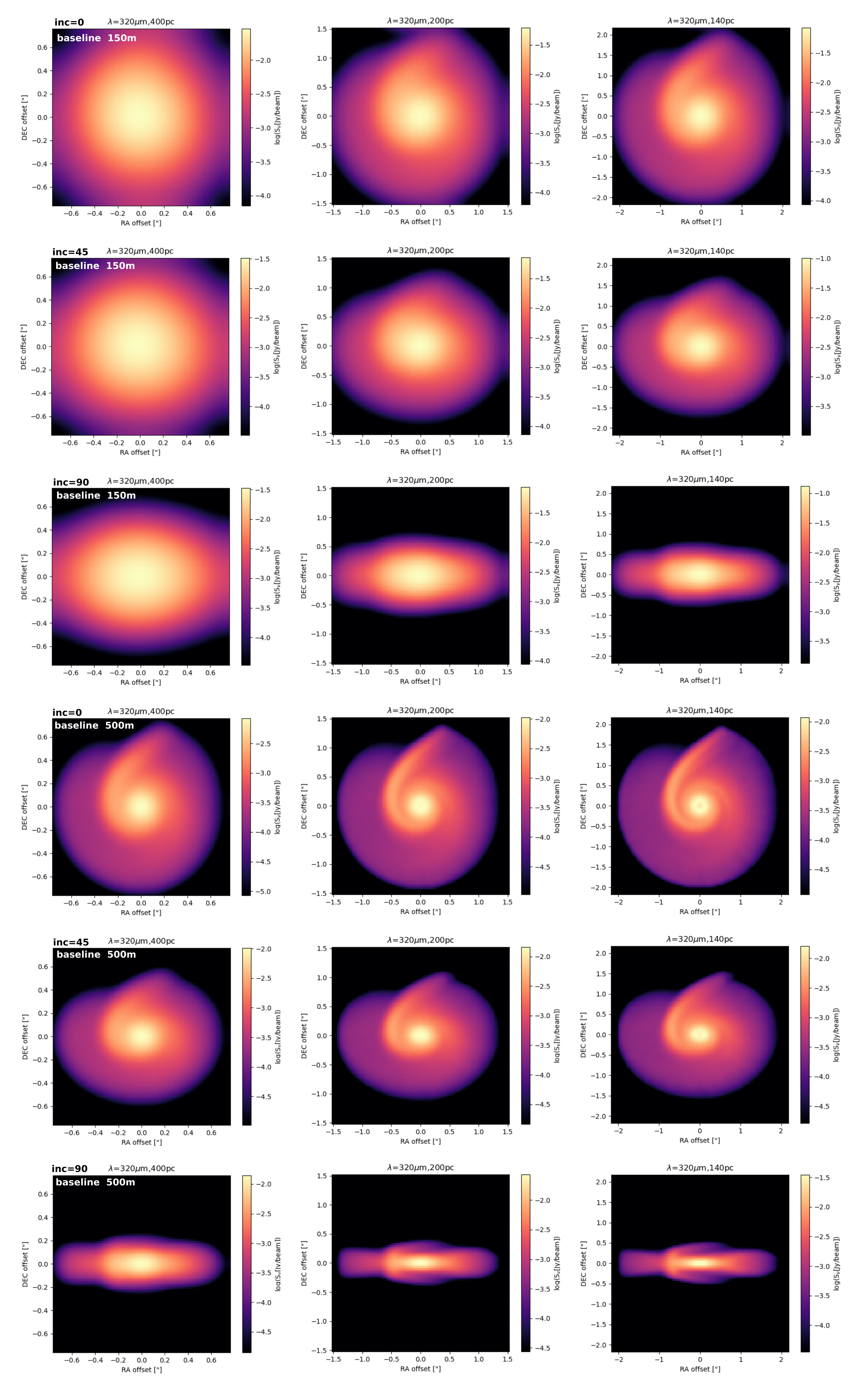}
\caption{Convolved radiation maps for 320 $\mu m$. 
Shown are the results of applying convolution, considering a PSF of the ALMA array at 320 $\mu m$,
assuming distances of 140, 200, and 400 pc, respectively, and for l.o.s. inclination angles of 0, 45, and 90 degrees.
Here, the physical gas density at the inner disk radius is chosen as $\simeq 10^{-12} {\rm g\,cm}^{-3}$.
The convolved images are provided for two different baselines of the ALMA array telescope, 150m, and 500m, respectively. 
The color bar represents intensity in Jy/beam and is normalized based on the maximum and minimum intensity at the specific wavelength.}
\label{fig:rb2_convolved_ALMA}
\end{figure*}

\subsubsection{Larger gas density}

To explore the effects of different density normalizations, we also generated radiation maps for the binary star run by applying larger gas densities at the inner radius of the disk, specifically $\rho_i = 10^{-11} \,\rm g\,cm^{-3}$,
as shown in Figure~\ref{fig:rb2_simple_all_inc_den10to11} 
and Figure~\ref{fig:rb2_profile_all_inc10to11}, which we display in the appendix for brevity. 

Compared to the smaller gas density, distinct differences are observed in the resulting radiation maps applying $\rho_i = 10^{-11} \,\rm g\,cm^{-3}$.
As shown in Figure~\ref{fig:rb2_simple_all_inc_den10to11} (in the appendix) which has been provided by the {\em SDD} approach, the disc is not optically thin for wavelengths around $5\mu$m  and the radiation, received  by the observer, originates from relatively shallower layers of the disk.
The cross and fork shape radiation that exist in the lower density models and is produced by the scattered light from the surface layers of the disc and the spirals does not exist in the higher density model.
At wavelengths of $20$ and $25,\mu$m, the disk with a larger density is more opaque compared to the smaller density model and the dark region seen in the disk has expanded vertically. This effect is particularly evident at an inclination of $90^\circ$  degree, relative to the line of sight.

The disk wind from the surface is more prominent at $20$ and $25,\mu$m and originates from upper layers of the disk compared to the lower gas density case. 
At longer wavelengths, for example, $320\mu$m, the disk and the wind both are visible and bright enough because of the thermal emission of the disk and the base of the wind.

Additionally, we observe differences in the radiation maps when employing the {\em SND} approach for the larger gas density of $\rho_{\rm i} = 10^{-11} \,\rm g\,cm^{-3}$.
Regarding Figure~\ref{fig:rb2_profile_all_inc10to11} (in the appendix), the disk appears opaque at 5, 20, and 25 $\mu$m, with radiation originating from the upper layers of the disk and the base of the disk wind.
As discussed earlier, an increase in gas density allows 0.1 $\mu$m dust particles to reach higher altitudes and become more widely distributed throughout the disk wind. This effect is evident in the radiation maps shown in Figure~\ref{fig:rb2_profile_all_inc10to11} (in the appendix), where the disk wind becomes more pronounced at these wavelengths.

Additionally, some features within the wind, which were not visible at lower gas densities, are now visible.
At longer wavelengths, such as 320 $\mu$m, both the disk and the base of the disk wind appear bright, with thermal radiation being received from both. Furthermore, some features within the disk, spirals, and the disk wind are clearly visible at 320 micron. 

We would like to note that the spiky features in the smaller wavelengths and $90^\circ$ inclination are produced because of the high optical depth in the disk. In order to eliminate them, we had to increase the number of scattered photon by at least two order of magnitude that would become very costly. However, after convolution these features would be vanished~(see Fig.~\ref{fig:rb2_convolved_MIRI_10to11} in the appendix).
\subsection{Providing the radiation map for observation}
\label{prepare_map_for_telescope}
Understanding how a specific telescope observes a real astronomical object is crucial for distinguishing which features in a preliminary 
radiation map can be practically detected by different instruments.

The radiation map obtained in the previous section requires further improvement to generate an image corresponding to a specific instrument's 
capabilities, including the wavelength and distance involved.
This improvement is achieved by applying convolution.

Convolution models the impact of a telescope's optics and atmospheric conditions on the images of an object.
Convolution combines the true image of the object with the PSF, producing the observed image by effectively spreading the light from each 
point source according to the PSF.
Based on the above definition, we utilized some routines available in RADMC3D code (Radmc3dpy) and obtained the convolved images of the 
dust continuum radiation map at 5 and 320 $\mu m$ wavelengths.

We only present the convolved image of the radiation maps for the binary system, based on the {\em SND} approach.

These wavelengths correspond to typical wavelengths of the MIRI/JWST and ALMA instruments, respectively.

The point spread function (PSF) modeling in MIRI \citep{2024A&A...689A...5D} and ALMA is rather complex.
Here, in order to simplify our analysis, we approximate the PSF using a Gaussian profile for both instruments.

We use a FWHM of $\simeq$ 0.18 arc seconds for the 5-micron MIRI wavelength.
For 320 microns and ALMA,
we can achieve approximately a FWHM of 0.44 arc seconds for a 150-meter baseline,
and 0.13 arc seconds for a 500-meter baseline, respectively.

In Figure~\ref{fig:rb2_convolved_MIRI},
we display the resulting images after applying convolution by a Point Spread Function (PSF) of 
the MIRI  instrument on the JWST at 5 $\mu m$,
at distances of 140, 200, and 400\,pc, respectively, 
and for inclination angles of 0, 45 and 90 degrees to the line of sight.

We observe that at 5 $\mu m$ features such as spiral arms are visible. 
It is important to note that the color bar is normalized to the maximum flux of the disk-jet 
system at 5 $\mu m$ and at a specific distance.
At 140 pc, the contrast between the disk and the spiral arms is lower, 
making the spiral arm feature less distinguishable.
However, at 200 pc, the spiral arms become more pronounced and easier to distinguish compared to the closer distance of 140 pc.

We find that with the typical MIRI wavelength, the spiral arms appear as two lobes when the 
disk is edge-on.
On the other hand, for a line of sight below 90 degrees, the spiral arms appear as arc-like structures.

\begin{figure*}
\centering
 \includegraphics[width=16cm]{\figurepath/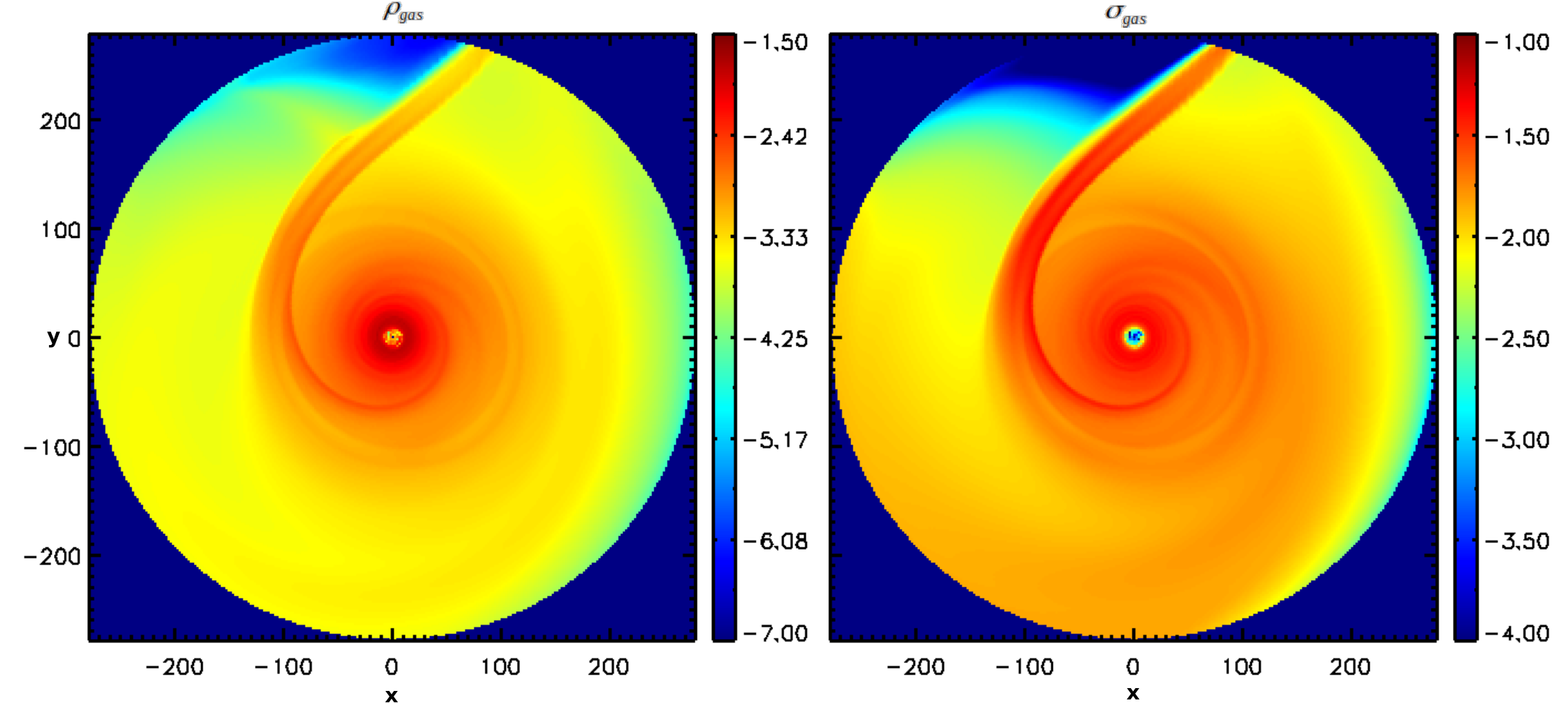}
\caption{Surface density. Shown are the snapshots of the gas density at the midplane  and the integrated gas density $\sigma_{gas}$ across the disk scale height for a binary system with the mass ratio of $q=0.2$ at time 6000 $t_{\rm i}$.}
\vspace{1cm}
\label{fig:surf_dens}
\end{figure*}

To resolve the spiral arms we may consider longer wavelengths as if we would observe 
with the ALMA array.
The wavelengths covered by ALMA range from 0.3 mm to 3.6 mm.
This wavelength range is crucial for studying cold and dusty regions of the universe, which are often invisible in optical and infrared wavelengths.
For example, \cite{2016Sci...353.1519P} used the ALMA Array to observe the protoplanetary disk around the young star Elias 2-27 at a wavelength of 1.3 mm. 
Their observations reveal two symmetric spiral arms located at a radius of 70 AU within the disk, 
at a distance of 140 pc.

Figure~\ref{fig:rb2_convolved_ALMA} illustrates the resulting images after applying convolution, considering an ALMA PSF at 320 $\mu m$, distances of 140, 200, and 400 pc, respectively, and for inclination angles of 0, 45 and 90 degrees.
The convolved images are provided for two different baselines of the ALMA array telescope, 150m, and 500m, respectively.

From Figure~\ref{fig:rb2_convolved_ALMA} we see that
when the source distance increases, thus the angular sizes decrease, it becomes more challenging to resolve features with a 150 m baseline. 
Specifically, at a distance of 400 \,pc the spiral structure is not visible with the 150 m baseline, 
but as we observe closer distances, an asymmetric structure begins to emerge clearly. 
At 140 pc, the substructures are resolved across all inclinations.

With a 500 m baseline, however, the resolution of disk-jet features improves significantly. The image corresponding to the 500 m baseline shows the asymmetric features in the disk and the upper layer of the disk and in the disk wind is much more pronounced.

Increasing the baseline in ALMA is equivalent to increasing the telescope aperture which results in higher resolution and smaller PSF.
Therefore, even at 400pc source distance, spiral arms in the accretion disk which are essentially
caused by tidal effects seem detectable with ALMA.

For the larger gas density, $\rho_{\rm i} = 10^{-11}\,\rm g\,cm^{-3}$, as figures~\ref{fig:10to11convolved} and \ref{fig:rb2_convolved_MIRI_10to11} (in the appendix) show, the disc and wind features can be better resolved in the convolved image at 320 $\mu$m  using a 500-meter baseline compared to smaller wavelengths. Furthermore, features such as spiral arms appear brighter at the base of the wind, and the vertical expansion of the opaque disk has increased compared to the model applying lower gas density. At 5~$\mu$m, no feature is observable unless the disk is edge-on, in which the spirals and some parts of the disk surface could be distingusihed by their different vertical extends.

\subsection{Radiation map vs gas density distribution}
Here, we discuss the differences we find between the dust continuum radiation map and the more simple picture that we derive from the dynamical simulation, such as gas density, or magnetic field distribution, which cannot actually be observed directly.

Comparing the density distribution, Figure~\ref{fig:2Dview_rho_rb2_q0.2}, and the radiation map, Figure~\ref{fig:rb2_profile_all_inc}, we find that the gas density map exhibits more small-size features, while the radiation map appears significantly smoother.
This difference naturally arises because the gas density is represented in slices, whereas the radiation maps are integrated along the line of sight.
So local features or substructure may be averaged out.

For a more reasonable comparison, we provide an integrated gas density along the disk's scale height in Figure~\ref{fig:surf_dens}, therefore illustrating the disk surface density, $\sigma_{\rm gas}$.
Here, the spiral arms are more pronounced, but the overall density distribution is smoother due to the integration or averaging applied.

If we consider a spiral arm in the density map that is inclined relative to the equatorial plane, then slicing through that arm may reveal only a single, faint feature. 
In contrast, integrating the radiation along the line of sight captures contributions from all layers, is effectively summing up all the slices.
This integration allows us to observe detailed structures in the spiral arm, such as a potential double spiral arm, that may not be visible in the pure density map.

In the radiative transfer simulations, we distribute dust in two distinct ways.
First, by placing it uniformly wherever gas exists, using a gas-to-dust ratio of 100 
({\em SDD} approach); 
and second, by placing it only in regions where the gas density is high enough to achieve 
a specific Stokes number for the dust particles ({\em SND} approach).

Applying the first kind of distribution, the dust is present throughout the gas, while in the second, it appears only in denser gas areas.
Consequently, the second approach allows the radiation maps to highlight the disk's gas features with a limited representation of the wind.

The radiation we receive from the overall system corresponds to regions with an optical depth of unity, which varies with wavelength.
For example, the radiation between 10 and 30 microns typically originates from the disk's upper layer, capturing surface features through light scattered by dust particles. 
In contrast, at longer wavelengths, such as 320 microns, thermal radiation from colder dust emerges, where the disk becomes optically thin, revealing deeper structures.

We note that in this study, we simulate only dust radiation to reflect gas density features. 
Predictions for gas velocity observations would require gas line radiation, which we defer to our future work.

\section{Conclusions}
In this paper, we have presented the results of 3D magnetohydrodynamic (MHD) simulations of a resistive, magnetized jet-launching 
circumstellar disk, which has been applied to both single star and binary star systems. 
We have employed the PLUTO code version 4.4.2, utilizing a unique combination of 3D MHD launching simulations and 3D radiative 
transfer to study the tidal effects caused by the companion star on the disk-jet structure.

Our results show that the jets are launched out of the accretion stream and are accelerated magneto-centrifugally.
 
As a major step forward, from these simulations, we have derived the first 3D radiation map of the dust continuum 
for such disk-jet structures, utilizing the radiative transfer code {"}RADMC-3D{"}.
In principle, this potentially allows us to compare MHD simulation results to observed disk-outflow features.
We summarize our main findings as follows:
 
(i) Our setup, utilizing full 3D spherical coordinates, operates consistently, and can establish
a consistent configuration of the dynamical disk-jet structure.
The jet material is initially launched out of the inner disk while launching gradually extends to larger radii over time. 
The simulation domain is sufficiently large to observe the launching of two distinct components; 
the faster outflow or jet originating from the inner disk region and the extended disk wind from larger radii. 
This confirms the continuous formation of the jet, consistent with our previous 3D studies 
\citep{2015ApJ...814..113S,2018ApJ...861...11S}, and earlier axisymmetric simulations in the literature. \\

(ii) Our results indicate that the implementation of the secondary star via a time-dependent Roche potential leads to clearly visible deviations from axial
symmetry in the disk, which in turn initiates the formation of spiral arms. 
This phenomenon is also observed in the jet, albeit with a delayed onset.
The density distribution clearly illustrates the growth and expansion of the spiral arms. 
As shown in our previous studies \citep{2018ApJ...861...11S,2022ApJ...925..161S}, 
these spiral arms exhibit rotational motion that is synchronized with the orbital motion of the secondary star.
Compared to our previous works, we were now able to consider significantly larger grid scales 
within the disk and jet.
Our domain spans approximately 1000 inner disk radii for the single star-disk simulation. 
Since the larger disk and thus a more substantial mass reservoir, the run time has been notably extended to about 1600 inner disk revolutions.
The adoption of spherical coordinates has also led to higher computational efficiency.\\

(iii) By comparing the outflow and accretion mass fluxes obtained from a similar control volume in single and binary star systems, we
find greater perturbations in the time evolution profiles of the mass fluxes in the binary star system. 
Additionally, we identify several peaks in these profiles, indicating that both the accretion and outflow mass fluxes are enhanced in a binary system. 
In conclusion, our findings confirm that the physical evolution of the ejected material is governed by the evolution
of the material that is accreted.
This finding further supports the discussion in \citep{2022ApJ...925..161S}, suggesting that angular momentum removal and tidal effects 
influence the disk-jet evolution more efficiently in a binary system.\\

(iv) We provide dust continuum radiation maps of the disk-jet structure based on the dust density profile obtained from two approaches.
In the  {\em SDD} approach, we only consider the small dust particles of size 0.1~$\mu$m which 
are well-coupled to the gas and follow the dust-to-gas density ratio of 0.01. 
Our findings reveal that spiral arms are visible in all of our un-convolved radiation maps, both in scattered light and thermal radiation, 
except for edge-on disks at 20 and 25 $\mu m$ wavelengths.
In the 5 and 320 $\mu m$  bands, where the disk is optically thin, spiral arms can be 
distinguished regardless of disk inclination. 
However, at 20 and 25 $\mu m$ where the disk becomes optically thick, these features are only distinguishable when the disk inclination is below 90 degrees.
We also detect the signature of spiral arms at the base of the disk wind.
Unlike the disk spiral arms, the outflow is only visible when the disk is viewed edge-on.\\

(v) In the {\em SND} approach, two dust particle sizes are considered and distributed based on their Stokes number.
The spiral arms are visible at 5, 20, and 25 microns, as well as at longer wavelengths. 
In this framework, even small dust grains primarily reside in the disk and lower layers of the disk wind, due to their large Stokes number 
resulting from the low gas density in the wind.
Thus, the wind becomes observable when small dust is coupled to the gas, while the spiral arms are detectable when the dust has settled.
Consequently, we primarily receive radiation from the lower part of the disk wind, while the majority of the radiation originates from the 
disk itself.
Therefore, observing the disk wind is more probable in young disks, where the dust is small and also not settled.\\

(vi) We have also presented radiation maps for binary system with an increased disk gas density
of $\rho_i = 10^{-11} \,\rm g\,cm^{-3}$ at the inner disk radius. 
Applying the {\em SDD} approach, the disk is not optically thin for wavelengths around $5\mu$m.
Thus, the radiation received by the observer originates from relatively shallower layers of the disk.
The cross and fork shape radiation that is visible in the lower density models arises from
scattered light from the surface layers of the disk. We do not observe a spirals pattern for
the higher-density model.
At wavelengths of $20$ and $25,\mu$m, the disk with a higher density is more opaque compared to the lower density model. 
The dark region seen in the disk has expanded vertically. 
However, the disk wind from the surface is more prominent at $20$ and $25,\mu$m. 
At longer wavelengths, e.g. $320\mu$m, the disk and the wind both are visible and bright enough 
because of the thermal emission of the disk and the base of the wind.\\

(vii) When applying the {\em SND} approach for the model with larger gas  density,
the disk appears opaque at 5, 20, and 25\,$\mu$m, with radiation originating from the upper 
layers of the disk and the base of the disk wind.
When increasing the disk gas density, the 0.1\,$\mu$m dust particles now reach higher altitudes 
and become more widely distributed throughout the disk wind.
Thus the disk wind becomes more pronounced at these wavelengths. 
Therefore, certain features within the wind, which were not visible at lower gas densities, now 
become visible. 
At longer wavelengths, such as 320\,$\mu$m, both the disk and the base of the disk wind appear 
bright, with thermal radiation being received from both. 
Furthermore, certain features within the disk, such as spirals, and the disk wind are clearly 
visible at 320 micron. 
\\

(viii) Applying a convolution that combines the true image of the object with the point spread function of the specific telescope, we were able to
capture more realistic images of the investigated systems.
 We only presented the convolved image of the radiation maps for the binary system, based on the {\em SND} approach.
We identified two wavelengths that yield well-convolved images with distinguishable features of the disk-jet structure, such as spiral arms.
The first wavelength is 5 $\mu m$, which stays in the observational range of the Mid-Infrared Instrument (MIRI), and the second is 320 $\mu m$, 
which is in the wavelength range of the ALMA array.
Additionally, considering image convolution, we observe that the spiral arms, when the disk is edge-on, should appear as two lobes with MIRI.
If the inclination angle relative to the line of sight is less than 90 degrees, the spiral arms would appear as arc-like structures. 
Furthermore, to improve the resolution of the spiral arms, we need to use longer wavelengths and the ALMA array. 
Depending on the ALMA baseline, we can achieve a full width at half maximum (FWHM) of approximately 0.44 arcseconds for a 150-meter baseline, 
and 0.13 arcseconds for a 500-meter baseline.
This resolution allows us to effectively resolve the spiral arms at all inclination angles relative to the line of sight.\\

In this paper, we have presented the first radiation maps of MHD jet launching assuming a typical dust distribution in the accretion disk-jet system, potentially observable to present-day infrared instrumentation.
Next future steps may include considering a time evolution of the dust distribution; longer, substantially more CPU-expensive simulations 
of the 3D binary system; and accounting for heating and cooling of the disk.


\begin{acknowledgements}
We dedicate this paper to the memory of the late Willy Kley, who passed away on Dec. 21, 2021 - far too early.
We acknowledge his continuous support, without which we would not be where we are now.
We express our gratitude to Andrea Mignone and the PLUTO team for providing us with access 
to their code.
We also extend our appreciation to Cornelis P.~Dullemond for allowing us to use the radiative 
transfer code RADMC3D.
Our simulations were carried out on the COBRA and RAVEN clusters of the Max Planck Society.

\end{acknowledgements}

\bibliographystyle{apj}
\bibliography{bibpaper20213d}

\begin{thebibliography}{63}
\expandafter\ifx\csname natexlab\endcsname\relax\def\natexlab#1{#1}\fi

\bibitem[{{Bajaj} {et~al.}(2024){Bajaj}, {Pascucci}, {Gorti}, {Alexander},
  {Sellek}, {Morrison}, {Gaspar}, {Clarke}, {Xie}, {Ballabio}, \&
  {Deng}}]{2024AJ....167..127B}
{Bajaj}, N.~S., {Pascucci}, I., {Gorti}, U., {Alexander}, R., {Sellek}, A.,
  {Morrison}, J., {Gaspar}, A., {Clarke}, C., {Xie}, C., {Ballabio}, G., \&
  {Deng}, D. 2024, \aj, 167, 127

\bibitem[{{Blandford} \& {Payne}(1982)}]{1982MNRAS.199..883B}
{Blandford}, R.~D. \& {Payne}, D.~G. 1982, \mnras, 199, 883

\bibitem[{Booth \& Clarke(2021)}]{10.1093/mnras/stab090}
Booth, R.~A. \& Clarke, C.~J. 2021, Monthly Notices of the Royal Astronomical
  Society, 502, 1569

\bibitem[{{Casse} \& {Keppens}(2002)}]{2002ApJ...581..988C}
{Casse}, F. \& {Keppens}, R. 2002, \apj, 581, 988

\bibitem[{{de Valon} {et~al.}(2022){de Valon}, {Dougados}, {Cabrit}, {Louvet},
  {Zapata}, \& {Mardones}}]{2022A&A...668A..78D}
{de Valon}, A., {Dougados}, C., {Cabrit}, S., {Louvet}, F., {Zapata}, L.~A., \&
  {Mardones}, D. 2022, \aap, 668, A78

\bibitem[{{Dicken} {et~al.}(2024){Dicken}, {Mar{\'\i}n}, {Shivaei}, {Guillard},
  {Libralato}, {Glasse}, {Gordon}, {Cossou}, {Kavanagh}, {Temim}, {Flagey},
  {Klaassen}, {Rieke}, {Wright}, {Alberts}, {Azzollini},
  {{\'A}lvarez-M{\'a}rquez}, {Bouchet}, {Bright}, {Cracraft}, {Coulais},
  {Detre}, {Engesser}, {Fox}, {Gaspar}, {Gastaud}, {Glauser}, {Hines},
  {Kendrew}, {Labiano}, {Lagage}, {Lee}, {Law}, {Morrison}, {Noriega-Crespo},
  {Jones}, {Patapis}, {Scheithauer}, {Sloan}, \& {Tamas}}]{2024A&A...689A...5D}
{Dicken}, D., {Mar{\'\i}n}, M.~G., {Shivaei}, I., {Guillard}, P., {Libralato},
  M., {Glasse}, A., {Gordon}, K.~D., {Cossou}, C., {Kavanagh}, P., {Temim}, T.,
  {Flagey}, N., {Klaassen}, P., {Rieke}, G.~H., {Wright}, G., {Alberts}, S.,
  {Azzollini}, R., {{\'A}lvarez-M{\'a}rquez}, J., {Bouchet}, P., {Bright}, S.,
  {Cracraft}, M., {Coulais}, A., {Detre}, O.~H., {Engesser}, M., {Fox}, O.~D.,
  {Gaspar}, A., {Gastaud}, R., {Glauser}, A.~M., {Hines}, D.~C., {Kendrew}, S.,
  {Labiano}, A., {Lagage}, P.-O., {Lee}, D., {Law}, D.~R., {Morrison}, J.~E.,
  {Noriega-Crespo}, A., {Jones}, O., {Patapis}, P., {Scheithauer}, S., {Sloan},
  G.~C., \& {Tamas}, L. 2024, \aap, 689, A5

\bibitem[{{Dorschner} {et~al.}(1995){Dorschner}, {Begemann}, {Henning},
  {Jaeger}, \& {Mutschke}}]{1995A&A...300..503D}
{Dorschner}, J., {Begemann}, B., {Henning}, T., {Jaeger}, C., \& {Mutschke}, H.
  1995, \aap, 300, 503

\bibitem[{{Dullemond} {et~al.}(2012){Dullemond}, {Juhasz}, {Pohl}, {Sereshti},
  {Shetty}, {Peters}, {Commercon}, \& {Flock}}]{2012ascl.soft02015D}
{Dullemond}, C.~P., {Juhasz}, A., {Pohl}, A., {Sereshti}, F., {Shetty}, R.,
  {Peters}, T., {Commercon}, B., \& {Flock}, M. 2012, {RADMC-3D: A
  multi-purpose radiative transfer tool}, Astrophysics Source Code Library,
  record ascl:1202.015

\bibitem[{{Fang} {et~al.}(2023){Fang}, {Wang}, {Herczeg}, {Hashimoto}, {Xu},
  {Nemer}, {Pascucci}, {Haffert}, \& {Aoyama}}]{2023NatAs...7..905F}
{Fang}, M., {Wang}, L., {Herczeg}, G.~J., {Hashimoto}, J., {Xu}, Z., {Nemer},
  A., {Pascucci}, I., {Haffert}, S.~Y., \& {Aoyama}, Y. 2023, Nature Astronomy,
  7, 905

\bibitem[{{Fendt}(2006)}]{2006ApJ...651..272F}
{Fendt}, C. 2006, \apj, 651, 272

\bibitem[{{Fendt} \& {{\v C}emelji{\'c}}(2002)}]{2002A&A...395.1045F}
{Fendt}, C. \& {{\v C}emelji{\'c}}, M. 2002, \aap, 395, 1045

\bibitem[{{Ferreira}(1997)}]{1997A&A...319..340F}
{Ferreira}, J. 1997, \aap, 319, 340

\bibitem[{{Ferreira}(2003)}]{2003astro.ph.11621F}
---. 2003, ArXiv Astrophysics e-prints

\bibitem[{{Flock} {et~al.}(2011){Flock}, {Dzyurkevich}, {Klahr}, {Turner}, \&
  {Henning}}]{2011ApJ...735..122F}
{Flock}, M., {Dzyurkevich}, N., {Klahr}, H., {Turner}, N.~J., \& {Henning}, T.
  2011, \apj, 735, 122

\bibitem[{{Flores-Rivera} {et~al.}(2023){Flores-Rivera}, {Flock}, {Kurtovic},
  {Husemann}, {Banzatti}, {Ringqvist}, {Kamann}, {M{\"u}ller}, {Fendt},
  {Garc{\'\i}a Lopez}, {Marleau}, {Henning}, {Carrasco-Gonz{\'a}lez}, {van
  Boekel}, {Keppler}, {Launhardt}, \& {Aoyama}}]{2023A&A...670A.126F}
{Flores-Rivera}, L., {Flock}, M., {Kurtovic}, N.~T., {Husemann}, B.,
  {Banzatti}, A., {Ringqvist}, S.~C., {Kamann}, S., {M{\"u}ller}, A., {Fendt},
  C., {Garc{\'\i}a Lopez}, R., {Marleau}, G.-D., {Henning}, T.,
  {Carrasco-Gonz{\'a}lez}, C., {van Boekel}, R., {Keppler}, M., {Launhardt},
  R., \& {Aoyama}, Y. 2023, \aap, 670, A126

\bibitem[{{Frank} {et~al.}(1999){Frank}, {Gardiner}, {Delemarter}, {Lery}, \&
  {Betti}}]{1999ApJ...524..947F}
{Frank}, A., {Gardiner}, T.~A., {Delemarter}, G., {Lery}, T., \& {Betti}, R.
  1999, \apj, 524, 947

\bibitem[{{Gueth} \& {Guilloteau}(1999)}]{1999A&A...343..571G}
{Gueth}, F. \& {Guilloteau}, S. 1999, \aap, 343, 571

\bibitem[{{Heese} {et~al.}(2017){Heese}, {Wolf}, {Dutrey}, \&
  {Guilloteau}}]{2017A&A...604A...5H}
{Heese}, S., {Wolf}, S., {Dutrey}, A., \& {Guilloteau}, S. 2017, \aap, 604, A5

\bibitem[{{Jaeger} {et~al.}(1994){Jaeger}, {Mutschke}, {Begemann}, {Dorschner},
  \& {Henning}}]{1994A&A...292..641J}
{Jaeger}, C., {Mutschke}, H., {Begemann}, B., {Dorschner}, J., \& {Henning}, T.
  1994, \aap, 292, 641

\bibitem[{{Krasnopolsky} {et~al.}(1999){Krasnopolsky}, {Li}, \&
  {Blandford}}]{1999ApJ...526..631K}
{Krasnopolsky}, R., {Li}, Z., \& {Blandford}, R. 1999, \apj, 526, 631

\bibitem[{{Kudoh} {et~al.}(1998){Kudoh}, {Matsumoto}, \&
  {Shibata}}]{1998ApJ...508..186K}
{Kudoh}, T., {Matsumoto}, R., \& {Shibata}, K. 1998, \apj, 508, 186

\bibitem[{{Labdon} {et~al.}(2023){Labdon}, {Kraus}, {Davies}, {Kreplin},
  {Zarrilli}, {Monnier}, {Le Bouquin}, {Anugu}, {Setterholm}, {Gardner},
  {Ennis}, {Lanthermann}, {ten Brummelaar}, {Schaefer}, \&
  {Harries}}]{2023A&A...678A...6L}
{Labdon}, A., {Kraus}, S., {Davies}, C.~L., {Kreplin}, A., {Zarrilli}, S.,
  {Monnier}, J.~D., {Le Bouquin}, J.-B., {Anugu}, N., {Setterholm}, B.,
  {Gardner}, T., {Ennis}, J., {Lanthermann}, C., {ten Brummelaar}, T.,
  {Schaefer}, G., \& {Harries}, T.~J. 2023, \aap, 678, A6

\bibitem[{{Larwood} {et~al.}(1996){Larwood}, {Nelson}, {Papaloizou}, \&
  {Terquem}}]{1996MNRAS.282..597L}
{Larwood}, J.~D., {Nelson}, R.~P., {Papaloizou}, J.~C.~B., \& {Terquem}, C.
  1996, \mnras, 282, 597

\bibitem[{{Lee} {et~al.}(2001){Lee}, {Stone}, {Ostriker}, \&
  {Mundy}}]{2001ApJ...557..429L}
{Lee}, C.-F., {Stone}, J.~M., {Ostriker}, E.~C., \& {Mundy}, L.~G. 2001, \apj,
  557, 429

\bibitem[{{Li}(1995)}]{1995ApJ...444..848L}
{Li}, Z. 1995, \apj, 444, 848

\bibitem[{{Long} {et~al.}(2008){Long}, {Romanova}, \&
  {Lovelace}}]{2008MNRAS.386.1274L}
{Long}, M., {Romanova}, M.~M., \& {Lovelace}, R.~V.~E. 2008, \mnras, 386, 1274

\bibitem[{{Mignone} {et~al.}(2007){Mignone}, {Bodo}, {Massaglia}, {Matsakos},
  {Tesileanu}, {Zanni}, \& {Ferrari}}]{2007ApJS..170..228M}
{Mignone}, A., {Bodo}, G., {Massaglia}, S., {Matsakos}, T., {Tesileanu}, O.,
  {Zanni}, C., \& {Ferrari}, A. 2007, \apjs, 170, 228

\bibitem[{{Mignone} {et~al.}(2012){Mignone}, {Zanni}, {Tzeferacos}, {van
  Straalen}, {Colella}, \& {Bodo}}]{2012ApJS..198....7M}
{Mignone}, A., {Zanni}, C., {Tzeferacos}, P., {van Straalen}, B., {Colella},
  P., \& {Bodo}, G. 2012, \apjs, 198, 7

\bibitem[{{Murphy} {et~al.}(2010){Murphy}, {Ferreira}, \&
  {Zanni}}]{2010A&A...512A..82M}
{Murphy}, G.~C., {Ferreira}, J., \& {Zanni}, C. 2010, \aap, 512, A82+

\bibitem[{{Nazari} {et~al.}(2024){Nazari}, {Tabone}, {Ahmadi}, {Cabrit}, {van
  Dishoeck}, {Codella}, {Ferreira}, {Podio}, {Tychoniec}, \& {van
  Gelder}}]{2024A&A...686A.201N}
{Nazari}, P., {Tabone}, B., {Ahmadi}, A., {Cabrit}, S., {van Dishoeck}, E.~F.,
  {Codella}, C., {Ferreira}, J., {Podio}, L., {Tychoniec}, {\L}., \& {van
  Gelder}, M.~L. 2024, \aap, 686, A201

\bibitem[{{Ouyed} {et~al.}(2003){Ouyed}, {Clarke}, \&
  {Pudritz}}]{2003ApJ...582..292O}
{Ouyed}, R., {Clarke}, D.~A., \& {Pudritz}, R.~E. 2003, \apj, 582, 292

\bibitem[{{Ouyed} \& {Pudritz}(1997)}]{1997ApJ...482..712O}
{Ouyed}, R. \& {Pudritz}, R.~E. 1997, \apj, 482, 712

\bibitem[{{Pascucci} {et~al.}(2023){Pascucci}, {Cabrit}, {Edwards}, {Gorti},
  {Gressel}, \& {Suzuki}}]{2023ASPC..534..567P}
{Pascucci}, I., {Cabrit}, S., {Edwards}, S., {Gorti}, U., {Gressel}, O., \&
  {Suzuki}, T.~K. 2023, in Astronomical Society of the Pacific Conference
  Series, Vol. 534, Protostars and Planets VII, ed. S.~{Inutsuka}, Y.~{Aikawa},
  T.~{Muto}, K.~{Tomida}, \& M.~{Tamura}, 567

\bibitem[{{P{\'e}rez} {et~al.}(2016){P{\'e}rez}, {Carpenter}, {Andrews},
  {Ricci}, {Isella}, {Linz}, {Sargent}, {Wilner}, {Henning}, {Deller},
  {Chandler}, {Dullemond}, {Lazio}, {Menten}, {Corder}, {Storm}, {Testi},
  {Tazzari}, {Kwon}, {Calvet}, {Greaves}, {Harris}, \&
  {Mundy}}]{2016Sci...353.1519P}
{P{\'e}rez}, L.~M., {Carpenter}, J.~M., {Andrews}, S.~M., {Ricci}, L.,
  {Isella}, A., {Linz}, H., {Sargent}, A.~I., {Wilner}, D.~J., {Henning}, T.,
  {Deller}, A.~T., {Chandler}, C.~J., {Dullemond}, C.~P., {Lazio}, J.,
  {Menten}, K.~M., {Corder}, S.~A., {Storm}, S., {Testi}, L., {Tazzari}, M.,
  {Kwon}, W., {Calvet}, N., {Greaves}, J.~S., {Harris}, R.~J., \& {Mundy},
  L.~G. 2016, Science, 353, 1519

\bibitem[{{Pesenti} {et~al.}(2003){Pesenti}, {Dougados}, {Cabrit}, {O'Brien},
  {Garcia}, \& {Ferreira}}]{2003A&A...410..155P}
{Pesenti}, N., {Dougados}, C., {Cabrit}, S., {O'Brien}, D., {Garcia}, P., \&
  {Ferreira}, J. 2003, \aap, 410, 155

\bibitem[{{Porth} \& {Fendt}(2010)}]{2010ApJ...709.1100P}
{Porth}, O. \& {Fendt}, C. 2010, \apj, 709, 1100

\bibitem[{{Pudritz} \& {Norman}(1983)}]{1983ApJ...274..677P}
{Pudritz}, R.~E. \& {Norman}, C.~A. 1983, \apj, 274, 677

\bibitem[{{Pudritz} {et~al.}(2007){Pudritz}, {Ouyed}, {Fendt}, \&
  {Brandenburg}}]{2007prpl.conf..277P}
{Pudritz}, R.~E., {Ouyed}, R., {Fendt}, C., \& {Brandenburg}, A. 2007,
  Protostars and Planets V, 277

\bibitem[{Rabenanahary {et~al.}(2022)Rabenanahary, Cabrit, Meliani, \& des
  For{\^e}ts}]{Rabenanahary2022}
Rabenanahary, M., Cabrit, S., Meliani, Z., \& des For{\^e}ts, G.~P. 2022,
  Astronomy \& Astrophysics

\bibitem[{{Rodenkirch} \& {Dullemond}(2022)}]{2022A&A...659A..42R}
{Rodenkirch}, P.~J. \& {Dullemond}, C.~P. 2022, \aap, 659, A42

\bibitem[{{Romanova} {et~al.}(2013){Romanova}, {Ustyugova}, {Koldoba}, \&
  {Lovelace}}]{2013MNRAS.430..699R}
{Romanova}, M.~M., {Ustyugova}, G.~V., {Koldoba}, A.~V., \& {Lovelace},
  R.~V.~E. 2013, \mnras, 430, 699

\bibitem[{{Shakura} \& {Sunyaev}(1973)}]{1973shakuraetal}
{Shakura}, N.~I. \& {Sunyaev}, R.~A. 1973, \aap, 24, 337

\bibitem[{{Sheikhnezami} \& {Fendt}(2015)}]{2015ApJ...814..113S}
{Sheikhnezami}, S. \& {Fendt}, C. 2015, \apj, 814, 113

\bibitem[{{Sheikhnezami} \& {Fendt}(2018)}]{2018ApJ...861...11S}
---. 2018, \apj, 861, 11

\bibitem[{{Sheikhnezami} \& {Fendt}(2022)}]{2022ApJ...925..161S}
---. 2022, \apj, 925, 161

\bibitem[{{Sheikhnezami} {et~al.}(2012){Sheikhnezami}, {Fendt}, {Porth},
  {Vaidya}, \& {Ghanbari}}]{2012ApJ...757...65S}
{Sheikhnezami}, S., {Fendt}, C., {Porth}, O., {Vaidya}, B., \& {Ghanbari}, J.
  2012, \apj, 757, 65

\bibitem[{{Sheikhnezami} \& {Sepahvand}(2024)}]{2024ApJ...966...82S}
{Sheikhnezami}, S. \& {Sepahvand}, M. 2024, \apj, 966, 82

\bibitem[{{Shibata} \& {Uchida}(1985)}]{1985PASJ...37...31S}
{Shibata}, K. \& {Uchida}, Y. 1985, \pasj, 37, 31

\bibitem[{{Solf} \& {Boehm}(1993)}]{1993ApJ...410L..31S}
{Solf}, J. \& {Boehm}, K.~H. 1993, \apjl, 410, L31

\bibitem[{{Somigliana} {et~al.}(2023){Somigliana}, {Testi}, {Rosotti}, {Toci},
  {Lodato}, {Tabone}, {Manara}, \& {Tazzari}}]{2023ApJ...954L..13S}
{Somigliana}, A., {Testi}, L., {Rosotti}, G., {Toci}, C., {Lodato}, G.,
  {Tabone}, B., {Manara}, C.~F., \& {Tazzari}, M. 2023, \apjl, 954, L13

\bibitem[{{Stepanovs} \& {Fendt}(2014)}]{2014ApJ...793...31S}
{Stepanovs}, D. \& {Fendt}, C. 2014, \apj, 793, 31

\bibitem[{{Stone} {et~al.}(1996){Stone}, {Hawley}, {Gammie}, \&
  {Balbus}}]{1996ApJ...463..656S}
{Stone}, J.~M., {Hawley}, J.~F., {Gammie}, C.~F., \& {Balbus}, S.~A. 1996,
  \apj, 463, 656

\bibitem[{{Tabone} {et~al.}(2022){Tabone}, {Rosotti}, {Lodato}, {Armitage},
  {Cridland}, \& {van Dishoeck}}]{2022MNRAS.512L..74T}
{Tabone}, B., {Rosotti}, G.~P., {Lodato}, G., {Armitage}, P.~J., {Cridland},
  A.~J., \& {van Dishoeck}, E.~F. 2022, \mnras, 512, L74

\bibitem[{{Tambovtseva} \& {Grinin}(2008)}]{2008AstL...34..231T}
{Tambovtseva}, L.~V. \& {Grinin}, V.~P. 2008, Astronomy Letters, 34, 231

\bibitem[{{Tzeferacos} {et~al.}(2013){Tzeferacos}, {Ferrari}, {Mignone},
  {Zanni}, {Bodo}, \& {Massaglia}}]{2013MNRAS.428.3151T}
{Tzeferacos}, P., {Ferrari}, A., {Mignone}, A., {Zanni}, C., {Bodo}, G., \&
  {Massaglia}, S. 2013, \mnras, 428, 3151

\bibitem[{{Uchida} \& {Shibata}(1985{\natexlab{a}})}]{1985PASJ...37..515U}
{Uchida}, Y. \& {Shibata}, K. 1985{\natexlab{a}}, \pasj, 37, 515

\bibitem[{{Uchida} \& {Shibata}(1985{\natexlab{b}})}]{Uchida1985}
---. 1985{\natexlab{b}}, \pasj, 37, 515

\bibitem[{{Ustyugova} {et~al.}(1995){Ustyugova}, {Koldoba}, {Romanova},
  {Chechetkin}, \& {Lovelace}}]{1995ApJ...439L..39U}
{Ustyugova}, G.~V., {Koldoba}, A.~V., {Romanova}, M.~M., {Chechetkin}, V.~M.,
  \& {Lovelace}, R.~V.~E. 1995, \apjl, 439, L39

\bibitem[{{Vaidya} {et~al.}(2011){Vaidya}, {Fendt}, {Beuther}, \&
  {Porth}}]{2011ApJ...742...56V}
{Vaidya}, B., {Fendt}, C., {Beuther}, H., \& {Porth}, O. 2011, \apj, 742, 56

\bibitem[{{Valeg{\r{a}}rd} {et~al.}(2022){Valeg{\r{a}}rd}, {Ginski}, {Dominik},
  {Bae}, {Benisty}, {Birnstiel}, {Facchini}, {Garufi}, {Hogerheijde}, {van
  Holstein}, {Langlois}, {Manara}, {Pinilla}, {Rab}, {Ribas}, {Waters}, \&
  {Williams}}]{2022A&A...668A..25V}
{Valeg{\r{a}}rd}, P.~G., {Ginski}, C., {Dominik}, C., {Bae}, J., {Benisty}, M.,
  {Birnstiel}, T., {Facchini}, S., {Garufi}, A., {Hogerheijde}, M., {van
  Holstein}, R.~G., {Langlois}, M., {Manara}, C.~F., {Pinilla}, P., {Rab}, C.,
  {Ribas}, {\'A}., {Waters}, L.~B.~F.~M., \& {Williams}, J. 2022, \aap, 668,
  A25

\bibitem[{{Wardle} \& {K\"onigl}(1993)}]{1993ApJ...410..218W}
{Wardle}, M. \& {K\"onigl}, A. 1993, \apj, 410, 218

\bibitem[{{Whelan} {et~al.}(2021){Whelan}, {Pascucci}, {Gorti}, {Edwards},
  {Alexander}, {Sterzik}, \& {Melo}}]{2021ApJ...913...43W}
{Whelan}, E.~T., {Pascucci}, I., {Gorti}, U., {Edwards}, S., {Alexander},
  R.~D., {Sterzik}, M.~F., \& {Melo}, C. 2021, \apj, 913, 43

\bibitem[{{Zanni} {et~al.}(2007){Zanni}, {Ferrari}, {Rosner}, {Bodo}, \&
  {Massaglia}}]{2007A&A...469..811Z}
{Zanni}, C., {Ferrari}, A., {Rosner}, R., {Bodo}, G., \& {Massaglia}, S. 2007,
  \aap, 469, 811

\end{thebibliography}

\appendix
\begin{figure*}
\centering
\includegraphics[width=18cm]{\figurepath/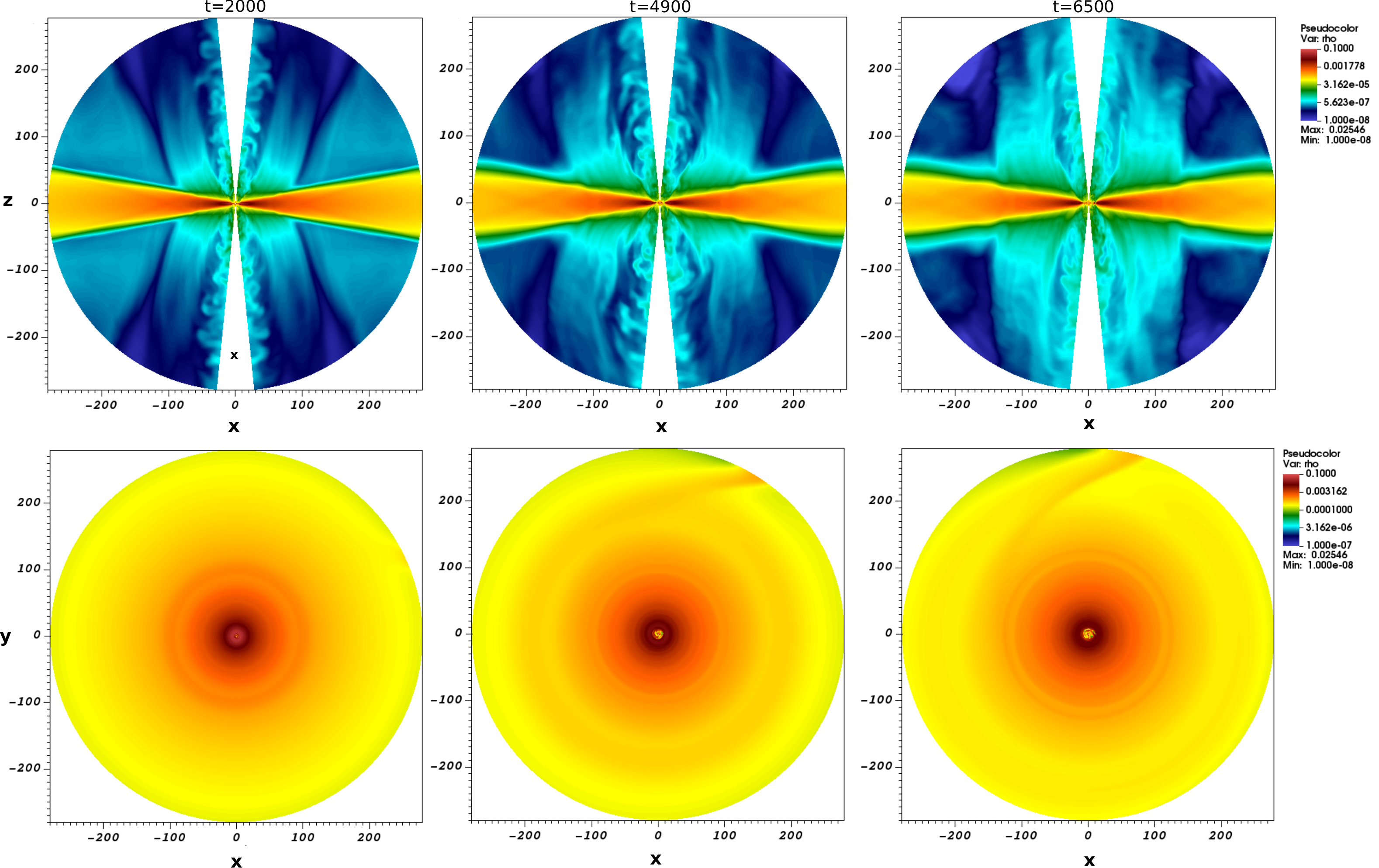}
\centering
\caption{Gas density distribution.
Shown are 2D snapshots of the gas density in our simulations for jetformation
in a binary star system (rb1), applying a mass ratio of $q=0.01$ of secondary to 
primary star. The stellar separation is $D=300$ within the $x-z$ plane.}
\label{fig:3Dview_rho_rb1_q0.01}
\end{figure*}

\begin{figure*}
\centering
\includegraphics[width=18cm]{\figurepath/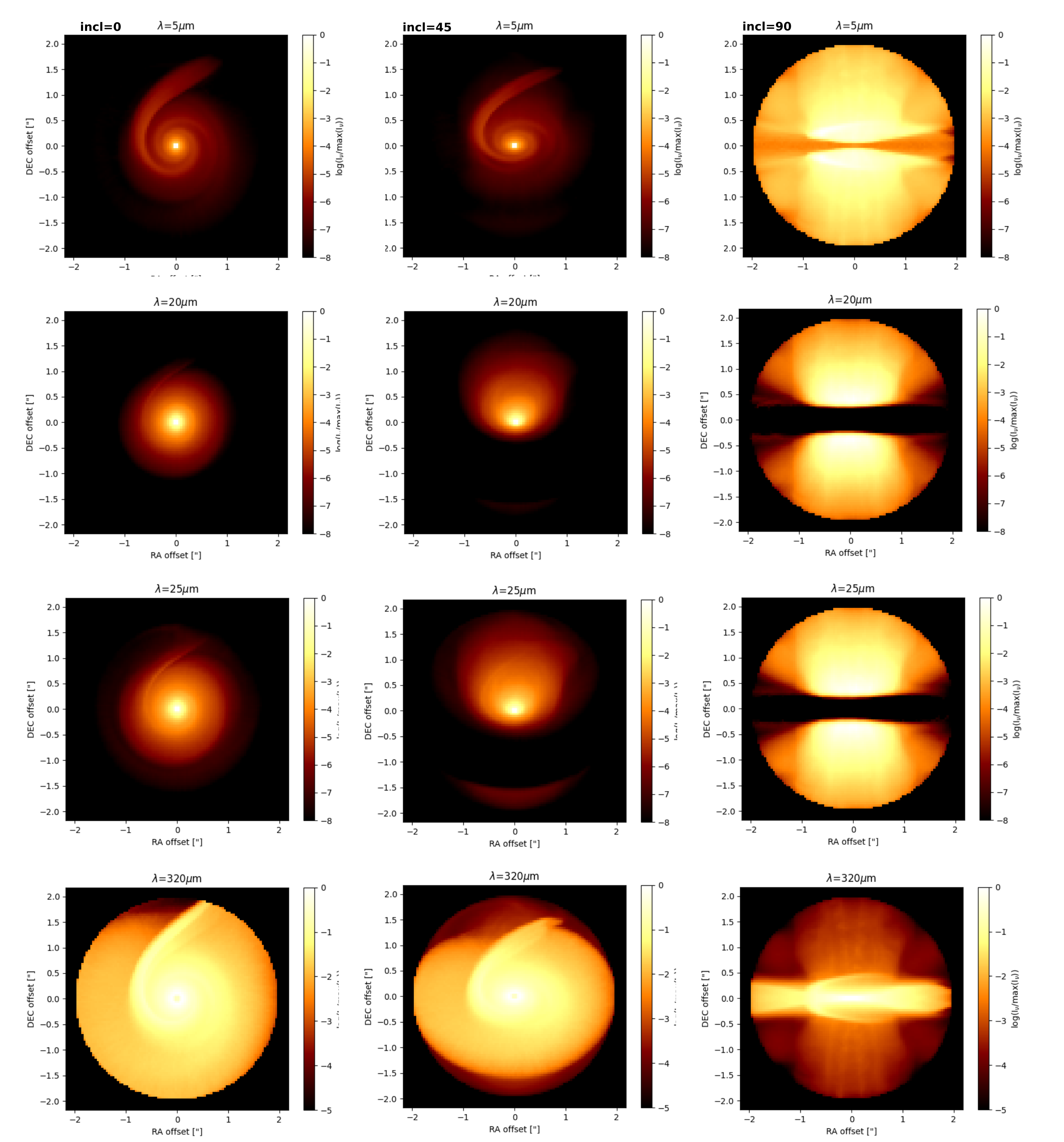}
\caption{Radiation maps for simulations considering a binary system.
Maps of the dust continuum radiation at different wavelengths are shown for the
simulation considering a binary star-disk-jet with a mass ratio of $q=0.2$ 
at dynamical time $t=6000$, and assuming a distance of 140\,pc. 
We apply different inclination angles to the l.o.s., such as $0, 45$ and $90$ degrees 
(from left to right).
Here, we have applied a one order of magnitude larger gas density at the inner disk 
radius, thus adopting $\rho_{\rm i}=10^{-11} \rm gr \,cm^{-3}$.
The color bar represents intensities in Jy/beam and is normalized considering
the maximum and minimum intensity at the specific wavelength.}
\label{fig:rb2_simple_all_inc_den10to11}
\end{figure*}
\begin{figure*}
\includegraphics[width=18cm]{\figurepath/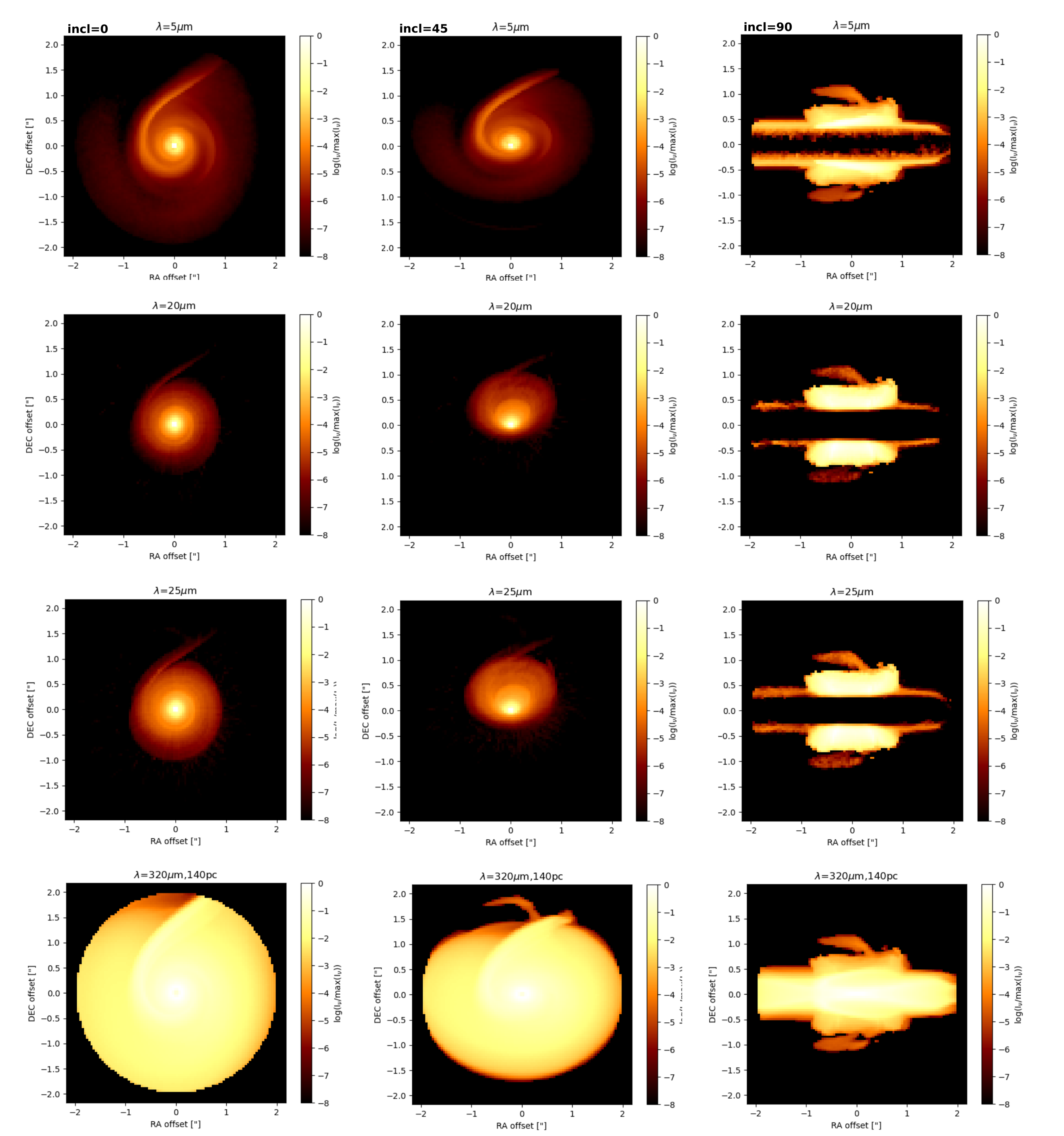}
\caption{ Radiation maps for simulations considering a binary system.
Maps of the dust continuum radiation at different wavelengths are shown for the
simulation considering a binary star-disk-jet with a mass ratio of $q=0.2$ 
at dynamical time $t=6000$, assuming a distance of 140\,pc.
Here, the density profile is obtained according to the Stokes number for
two dust particle sizes, specifically $10^{-5}$ and $10^{-4}$ cm. 
Here, we have applied a one order of magnitude larger gas density at the inner disk 
radius, thus adopting $\rho_{\rm i}=10^{-11} \rm gr \,cm^{-3}$.
We apply different inclination angles to the l.o.s., such as $0, 45$, and $90$ degrees
(from left to right).
The color bar represents intensities in Jy/beam and is normalized considering
the maximum and minimum intensity at the specific wavelength.}
\label{fig:rb2_profile_all_inc10to11}
\end{figure*}
\begin{figure*}
\centering
 \includegraphics[width=17cm]{\figurepath/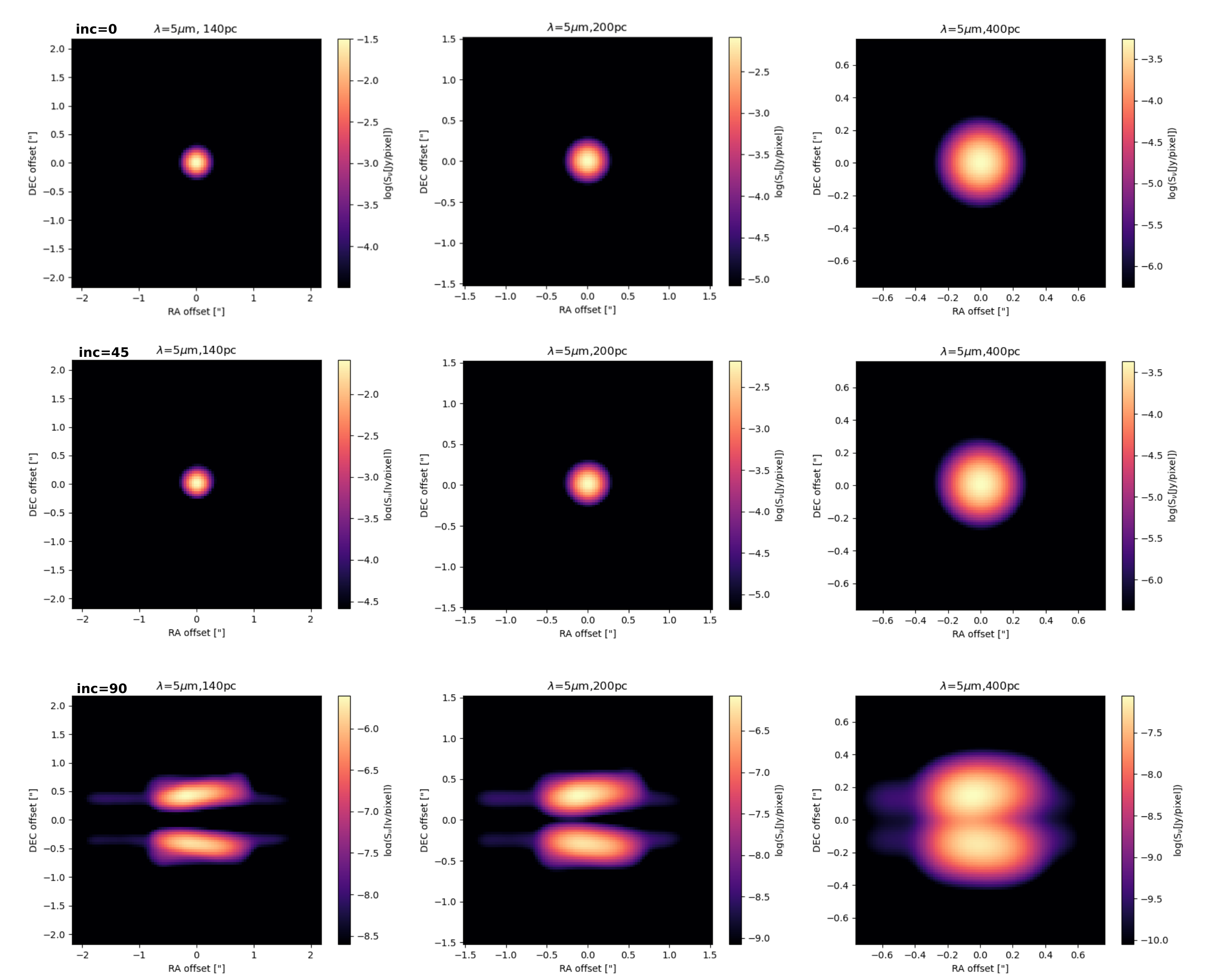}
\caption{Convolved radiation map for 5\,$\mu m$.
The image displays the results of applying a convolution by assuming a PSF and using the MIRI instrument on the JWST and for distances of 140, 200, and 400 pc, respectively, and for l.o.s. inclination angles of 0, 45, and 90 degrees.
Here, we have applied a one order of magnitude larger gas density at the inner disk 
radius, thus adopting $\rho_{\rm i}=10^{-11} \rm gr \,cm^{-3}$.
The color bar represents intensity in Jy/beam and is normalized based on the maximum and minimum intensity 
at the specific wavelength.}
\vspace{1cm}
\label{fig:rb2_convolved_MIRI_10to11}
\end{figure*}

\begin{figure*}
\centering
\includegraphics[width=14.5cm]{\figurepath/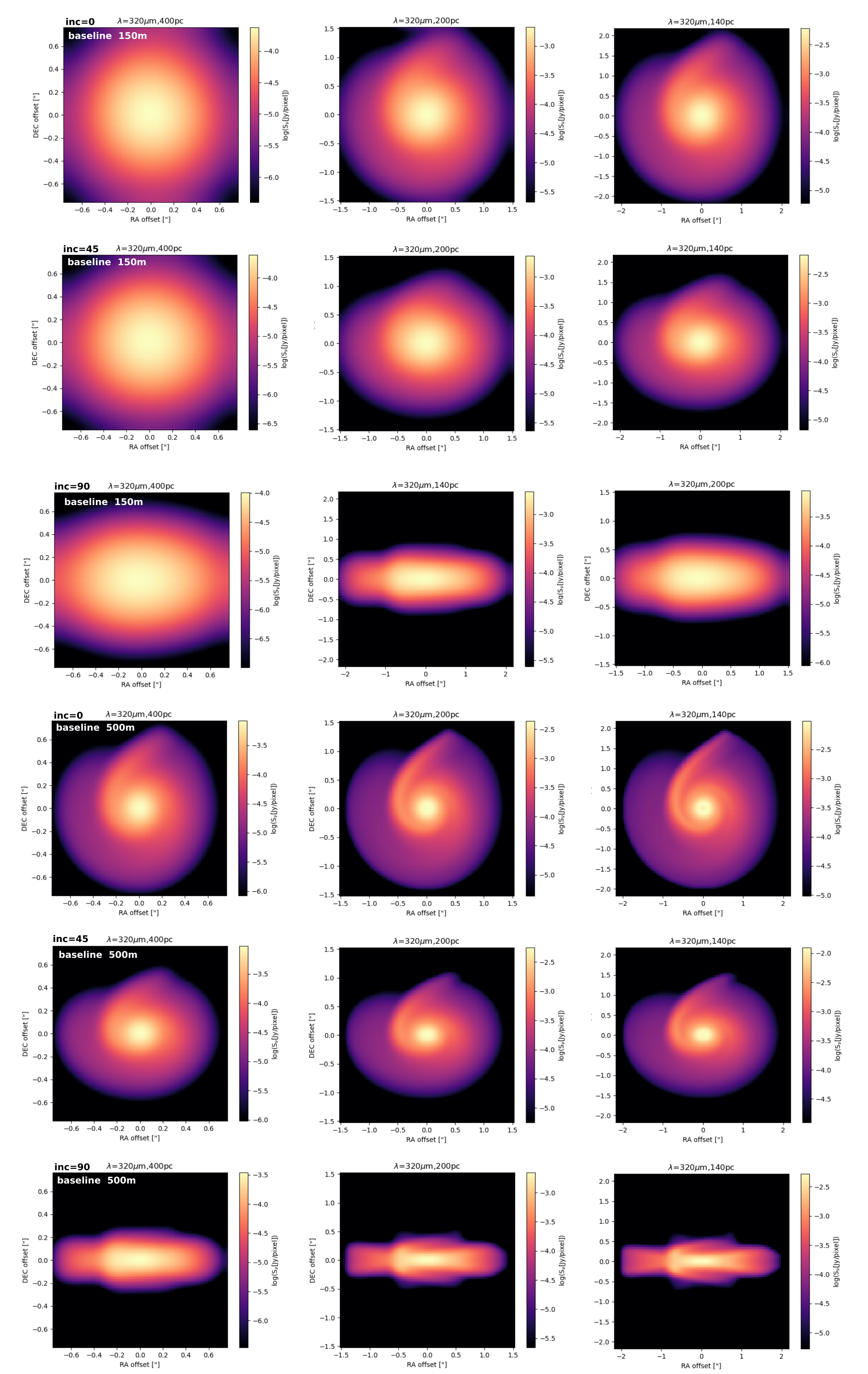}
\caption{Convolved radiation maps for 320\,$\mu m$.
Shown are the radiation maps derived above after applying convolution, 
considering an example PSF of the ALMA array at 320 $\mu m$, 
assuming distances of 140, 200, and 400 pc, respectively, 
and for l.o.s. inclination angles of 0, 45, and 90 degrees. 
Here, we have applied a one order of magnitude larger gas density at the inner disk 
radius, thus adopting $\rho_{\rm i}=10^{-11} \rm gr \,cm^{-3}$.
The convolved images are generated for two different baselines of the ALMA array,
that is 150m and 500m, respectively. 
The color bar represents intensity in Jy/beam and is normalized considering 
the maximum and minimum intensity at the specific wavelength.}
\label{fig:10to11convolved}
\end{figure*}

\end{document}